\documentclass[reprint,showpacs,aps,pra,
floatfix,
superscriptaddress
]{revtex4-1}

\usepackage[utf8]{inputenc}
\usepackage[T1]{fontenc}
\usepackage[]{graphicx}
\usepackage{color}
\usepackage{xcolor}
\usepackage{framed}
\usepackage[english]{babel}

\usepackage{soul}

\usepackage[bookmarks,bookmarksopen,bookmarksdepth=2]{hyperref}
\hypersetup{
    colorlinks=true,
    citecolor=blue,
    linkcolor=blue,
    filecolor=magenta,
    urlcolor=cyan,
}

\usepackage{url}

\usepackage{microtype}
\usepackage{braket}
\usepackage{esvect}
\usepackage[export]{adjustbox}

\usepackage{amsmath,amsfonts,amssymb}
\DeclareMathOperator{\tr}{tr}
\usepackage{mathtools}
\usepackage{bbm}

\usepackage[position=top,caption=false]{subfig}

\let\originaleqref\eqref
\renewcommand{\eqref}{Eq.~\originaleqref}

\newcommand{\trace}[1]{\tr\left(#1\right)} 
\newcommand{\mbf}[1]{\mathbf{#1}}

\newcommand{\correc}[1]{{#1}}
\newcommand{\BraKet}[2]{\left<#1|#2\right>}
\newcommand{\KetBra}[2]{|#1\rangle\langle#2|}

\begin{document}

\title{Scrambling and quantum chaos indicators from long-time properties of operator distributions}

\author{Sivaprasad Omanakuttan}
\email{somanakuttan@unm.edu}
\affiliation{Center for Quantum Information and Control, Department of Physics and Astronomy, University of New Mexico, Albuquerque, New Mexico 87131, USA}
\author{Karthik Chinni}
\email{karthik.chinni@polymtl.ca}
\affiliation{Department of Engineering Physics, \'Ecole Polytechnique de Montr\'eal, 2500 Chem. de Polytechnique, Montr\'eal, Quebec H3T 1J4, Canada}
\affiliation{Center for Quantum Information and Control, Department of Physics and Astronomy, University of New Mexico, Albuquerque, New Mexico 87131, USA}
\author{Philip Daniel Blocher}
\email{blocher@unm.edu}
\affiliation{Center for Quantum Information and Control, Department of Physics and Astronomy, University of New Mexico, Albuquerque, New Mexico 87131, USA}
\author{Pablo M. Poggi}
\email{ppoggi@unm.edu}
\affiliation{Center for Quantum Information and Control, Department of Physics and Astronomy, University of New Mexico, Albuquerque, New Mexico 87131, USA}

\date{\today}
\bigskip

\begin{abstract} 
Scrambling is a key concept in the analysis of nonequilibrium properties of quantum many-body systems. Most studies focus on its characterization via out-of-time-ordered correlation functions (OTOCs), particularly through the early-time decay of the OTOC. However, scrambling is a complex process which involves operator spreading and operator entanglement, and a full characterization requires one to access more refined information on the operator dynamics at several timescales. In this work we analyze operator scrambling by expanding the target operator in a complete basis and studying the structure of the expansion coefficients treated as a coarse-grained probability distribution in the space of operators. We study different features of this distribution, such as its mean, variance, and participation ratio, for the Ising model with longitudinal and transverse fields, kicked collective spin models, and random circuit models. We show that the long-time properties of the operator distribution display common features across these cases, and discuss how these properties can be used as a proxy for the onset of quantum chaos. Finally, we discuss the connection with OTOCs and analyze the cost of probing the operator distribution experimentally using these correlation functions.
\end{abstract}

\maketitle
\noindent

\section{Introduction}\label{section:introduction}

Scrambling refers to the spreading of initially localized information to the rest of the degrees of freedom in a many-body system \cite{Patrick_Hayden_2007,Yasuhiro_Sekino_2008,lashkari2013,hosur2016,swingle2016}. It plays an important role in describing diverse phenomena such as closed-system thermalization \cite{lewis2019,claeys2021}, dynamical phase transitions \cite{heyl2018,dag2019,wang2019}, sampling hardness in random quantum circuits \cite{boixo2018,bouland2019}, and information retrieval in black holes \cite{Yasuhiro_Sekino_2008,Patrick_Hayden_2007,basko2006metal}. One of the most prominent quantifiers of scrambling is the out-of-time-ordered correlator (OTOC), which is a four-point correlation function of the form $\langle W^\dagger(t) V^\dagger(0) W(t)V(0)\rangle$, together with the closely related square commutator $\langle \vert [W(t),V(0)] \vert^2\rangle$ \cite{Maldacena2016,cotler2017,swingle2018}.

OTOCs and scrambling have become important actors in the dynamical characterization of chaos in many-body quantum systems \cite{rozenbaum2017,garciamata2018,garciamata2022}. Quantum chaos is typically defined in terms of kinematic features like statistical properties of energy spectra and their connection to random matrix theory \cite{bohigas1984,Haake1991,wimberger2014,kos2018}. However, a unifying \textit{dynamical} description of chaos in general quantum systems remains an outstanding challenge. Many studies of scrambling in quantum systems have focused on understanding its early-time behavior, particularly through the decay of OTOCs, and have sought to define a quantum analogue of the Lyapunov exponent. Nevertheless, there are cases of quantum systems whose kinematic properties follow random matrix theory predictions (and are thus `quantum chaotic'), for which the OTOC does not decay exponentially \cite{fortes2019,Roberts2015diagnosing,aleiner2016microscopic,shenker2014black,shenker2015stringy}. Conversely, some quantum systems with classically integrable counterparts showing unstable fixed points can lead to exponential decay of OTOCs \cite{kidd2021saddle, Xu2020,pilatowsky2020}. 

More generally, scrambling is a complex process which can be described by the way an initially simple operator evolves into a complicated superposition of configurations belonging to an exponentially large operator space \cite{zhuang2019scrambling,Parker2019}. As such, it is bound to require methods to characterize it that go beyond the short-time behavior of a single correlation function. Indeed, recent studies on operator growth have developed a more thorough characterization of scrambling by analyzing the dynamics of operators in operator space, most notably using the Krylov representation \cite{Parker2019,noh2021}. Moreover, some studies have found important links between quantum chaos and the {long}-time behavior of the OTOC \cite{fortes2019}, rather than its initial decay. In this context it is useful to further probe the connection between the long-time properties of evolving operators and quantum chaos, and to explore other objects of interest beside the OTOC which allow us to construct diagnostics of scrambling in the long-time regime. An additional motivation is the inherent complexity of accessing OTOCs experimentally: even with the extraordinary control and isolation capabilities found in state-of-the-art quantum simulation experiments \cite{landsman2019,blok2021,mi2021}, accessing OTOCs require costly resources such as use of auxiliary systems or time-reversal operations \cite{garttner2017,PhysRevX.7.031011,swingle2016,arXiv.1607.01801}, among others, and thus require considerably more effort when compared to the usual quench-dynamics experiments (with notable exceptions, see \cite{vermersch2019,PhysRevLett.124.240505,blocher2022}).

In this work we purposely steer away from OTOCs and focus on studying scrambling in quantum systems by analyzing directly the dynamical properties of a suitably-defined probability distribution $\{P_k(t)\}$ over a coarse-grained operator basis. This distribution can be defined for arbitrary quantum systems, and here we focus on the cases of models of many spin-$1/2$ particles (including quantum circuits on qubits) and models of collective spins, which are effectively described by a single large spin $J$ \cite{lerose2020}. The dynamical transition from `simple' to `complex' operators is then encoded in different properties of the distribution such as its mean, variance, and (de)localization, which signify the growth and spreading of operators over the degrees of freedom of the system. We apply this framework to paradigmatic models of quantum chaos such as the `tilted'-field Ising model \cite{karthik2007entanglement} and the quantum kicked top \cite{Haake1987}, and show that analysis of the long-time averages of the distribution properties and their temporal fluctuations are good indicators of the onset of chaos in these systems. Furthermore, we discuss how some of these features can distinguish different properties of the nonergodic regimes, and show that both models can show very similar behavior in this picture even though their physical properties are quite different. We also apply these tools to the study of random quantum circuits \cite{fisher2022random,harrow2009random,oliveira2007,arute2019quantum} with a tunable number of $T$-gates, and analyze how the properties of scrambling change as just a small fraction of non-Clifford gates are included in the dynamics. Finally the connection between averages of OTOCs and the moments of the distribution $\{P_k(t)\}$ is analyzed and we discuss the number of OTOC measurements needed to reconstruct the different measures we study.

Our work extends previous studies that have focused on the properties of the operator distribution in the study of scrambling \cite{hosur2016,qi2019,Roberts2018,schuster2022}. Notably, these also include NMR experiments which routinely analyze the size of active clusters of spins, a quantity that is closely related to the mean operator size in the Heisenberg picture \cite{alvarez2010,dominguez2021}. It also complements the approach of Ref.~\cite{fortes2019}, which studied the connection between quantum chaos and the long-time properties of OTOCs by considering the properties of the operator distribution directly in a similar regime.

The rest of the work is organized as follows. 
In Sec.~\ref{sec:operator_evolution} we discuss scrambling for general quantum systems, and define the coarse-grained probability distribution as the object that characterizes it. 
In Sec.~\ref{sec:scrambling_and_chaos_in_the_tilted_field_Ising_model} the tilted-field Ising model is studied via numerical simulations, and we analyze the evolution of the probability distribution in its integrable and chaotic regimes.
In Sec.~\ref{sec:scrambling_and_chaos_in_the_quantum_kicked_top} we study the quantum kicked top, which is a collective spin model with a well-defined classical limit, and use it to relate the long-time properties of scrambling with the chaotic properties of the model. 
Then, in Sec.~\ref{sec:Scrambling in circuit model} we study a model of random Clifford circuits perturbed by a tunable number of $T$-gates, and study how the properties of the operator distribution change as the number of non-Clifford gates increases. Section~\ref{sec:otocs_and_connections} discusses the connection between the operator distribution $\{P_k(t)\}$ and averages of OTOCs. Finally, we give an outlook and discuss potential future work in Sec.~\ref{sec:conclusions_and_future_work}.

\section{Operator evolution and measures of scrambling}
\label{sec:operator_evolution}

Consider a quantum system on a finite-dimensional Hilbert space of dimension $d$, with evolution from $t=0$ to some arbitrary time $t$ described by an unitary operator $\hat{U}(t)$. We will focus on the dynamics of a generic hermitian operator $\hat{O}$, which can be expressed as
\begin{equation}
    \hat{O}(t)=\hat{U}^\dagger(t) \hat{O} \hat{U}(t) = \sum\limits_{j=0}^D f[\hat{\Lambda}_j;\hat{O}(t)] \hat{\Lambda}_j,
    \label{eqn:evol_O}
\end{equation}
where $\{\hat{\Lambda}_j\}$, $j=0,1,\ldots,d^2-1\equiv D$ is an operator basis which we take to be orthonormal, i.e.  $\trace{\hat{\Lambda}_i^\dagger \hat{\Lambda}_j}=\delta_{ij}$. We also take $\hat{\Lambda}_0 = \mathbb{I}/\sqrt{d}$ throughout, and consider $\tr \hat{O}=0$. The scalar coefficients $\{f[\hat{\Lambda}_j,\hat{O}(t)]\}$ describe the dynamics of $\hat{O}(t)$ in the chosen basis, and at all times satisfy the normalization condition 
\begin{equation}
    \sum\limits_j \left\lvert f[\hat{\Lambda}_j,\hat{O}(t)] \right\rvert^2  = \trace{\hat{O}(t)^2} =  \trace{\hat{O}^2}
    \label{eqn:norm_O}
\end{equation}

\begin{figure}[t!]
\centering{\includegraphics[width=0.9\linewidth]{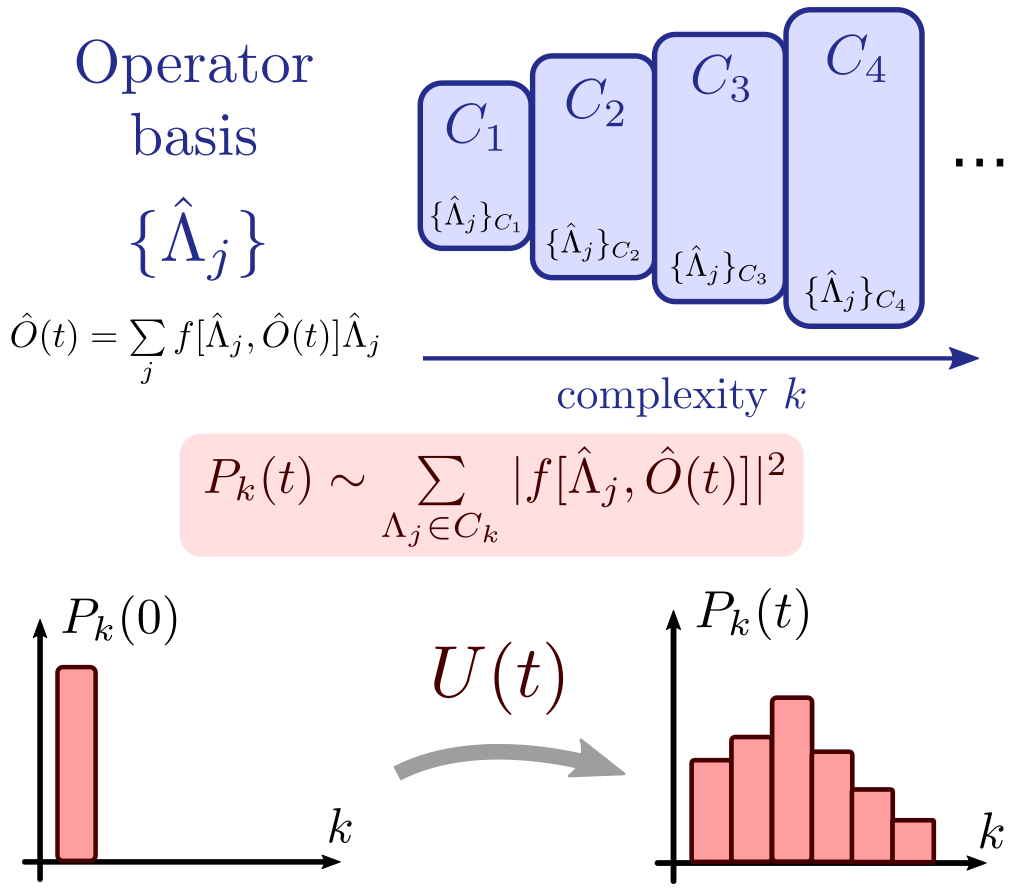}}
\caption{Schematic picture of scrambling diagnosed via a coarse-grained operator distribution. An operator basis, typically containing exponentially many elements, is divided into sets with common characteristics $C_1$, $C_2$, etc. A typical example of this is the size or weight of a multibody Pauli operator in the case of spin-$1/2$ particles. From this grouping, a probability distribution is defined for the evolution of an operator $\hat{O}(t)$. At $t=0$, the distribution is localized at low complexity index. The scrambling process generated by $U(t)$ spreads the distribution and shifts it towards higher complexity.}
\label{fig:figure1}
\end{figure}

Equation~(\ref{eqn:norm_O}) allows us to treat the set of squared coefficient amplitudes as a probability distribution (after normalization) \footnote{
The coefficients $\{f[\hat{\Lambda}_j,\hat{O}(t)]\}$ can also be regarded as the complex amplitudes of an ``operator wavefunction'' \cite{Roberts2018}, and thus Eqn. (\ref{eqn:norm_O}) represents the normalization of such wavefunction.}. Additionally, in many situations of interest the operator basis admits a natural ordering related to some notion of complexity of the operator, which we schematically depict in Fig. \ref{fig:figure1}. Qualitatively, we consider this ordering to induce a natural grouping of the basis, which has the form 
\begin{equation}
    \{\hat{\Lambda}_j\} \rightarrow \{\{\hat{\Lambda}_j\}_{C_1}, \{\hat{\Lambda}_j\}_{C_2},\ldots,\{\hat{\Lambda}_j\}_{C_k},\ldots,\{\hat{\Lambda}_j\}_{C_{k_\mathrm{max}}} \}
    \label{eqn:basis_group}
\end{equation}
where the complexity of operators is considered to grow as the index $k=1,2,\ldots,k_\mathrm{max}$ increases and $\sum_k \mathrm{dim}(C_k)=D$. For instance, this structure could correspond to a multibody Pauli basis where $k$ corresponds to the weight or size of each operator, as we will discuss in Sect. \ref{ssec:operator_evolution_pauli}. Leveraging this structure we define the following coarse-grained probability distribution for the operator $\hat{O}(t)$
\begin{equation}
    P_k(t) = \frac{1}{\trace{\hat{O}^2}} \sum\limits_{\Lambda_j \in C_k} \left\lvert f[\hat{\Lambda}_j,\hat{O}(t)] \right\rvert^2,
    \label{eqn:Pk_general}
\end{equation}
which naturally satisfies $\sum_k P_k(t) = 1$. We are interested in situations where the initial operator $\hat{O}$ is included within one of the groups of low complexity (say $k=1$), and our goal is to characterize the scrambling process in the system via the evolution of the distribution $P_k(t)$ over time, as depicted in Fig. \ref{fig:figure1}. To characterize the distribution we will consider the first two cumulants of the distribution, 
\begin{align}
    \mu(t) & = \overline{k(t)} = \sum\limits_{k=1} k P_k(t) \\
    \sigma(t) & = \sqrt{\overline{k(t)^2} - \mu(t)^2}
\end{align} 
as well as its inverse participation ratio (IPR), 
\begin{equation}
    \eta_{\mathrm{IPR}}(t) = \sum\limits_j P_j(t)^2.
    \label{eq:IPR_scrambling}
\end{equation}
Here $\eta_{\mathrm{IPR}}\simeq 1$ indicates a very localized distribution (``low participation''), while $\eta_{\mathrm{IPR}}\simeq 1/k_{\mathrm{max}}$ shows a delocalized distribution (``high participation''). The IPR has been extensively used to measure delocalization of eigenstates in quantum chaos \cite{Sieberer2019}, and a similar object has been used in the context of resource theories of Magic \cite{leone2022}. Here we will use it to assess the delocalization of the operator distribution in the scrambling process. We point out that higher-order cumulants of the full distribution in the Krylov basis have been studied recently in \cite{bhattacharjee2022cumulants}.

In Appendix \ref{appendix:haar} we show that for Haar-random evolution the operator $\hat{O}(t)$ will be uniformly spread in any operator basis. The resulting distribution thus has the form
\begin{equation}
    P_k \rightarrow \frac{\mathrm{dim}(C_k)}{d^2-1},
    \label{eqn:Pk_haar}
\end{equation}
and it is straightforward to compute the indicators introduced once the sets $\{C_k\}$ are defined. This limiting case will be helpful in order to study the onset of chaos as dictated by the randomization of the evolution. In the following subsections we investigate two cases of interest: systems of $N$ spin-$\frac{1}{2}$ particles or qubits, and collective spin systems described by a single large collective spin $J=N/2$.

\subsection{Systems of many spin-$1/2$ particles}
\label{ssec:operator_evolution_pauli}
Consider a system of $N$ spin-$1/2$ particles (with $d=2^N$) and the basis of multibody Pauli operators $\mathcal{P}^{\otimes N}$, where $\mathcal{P}=\{I,X,Y,Z\}=\{\mathbb{I},\sigma_x,\sigma_y,\sigma_z\}/\sqrt{2}$. This scenario encompasses many relevant models for the study of quantum chaos and scrambling, like the Ising model with a longitudinal and transverse field \cite{karthik2007entanglement}, random circuits on qubits \cite{fisher2022random,harrow2009random,oliveira2007}, and models spin chains with impurities or interactions beyond nearest-neighbors \cite{santos2011,santos2012,santos2020}. Each element of the Pauli basis can be assigned a size (or weight) $1\leq s(\hat{Q})\leq N$, which corresponds to the number of sites the operator acts non-trivially on. For instance, for $N=3$, $s(IXY)=2$, while $s(IZI)=1$. We will consider $s(\hat{Q})$ to be the measure of complexity of the basis elements in these systems. With this, the grouping of \eqref{eqn:basis_group} takes the form
\begin{equation}
    \{\hat{Q}_\mbf{j}\} \rightarrow \{ \{\hat{Q}_\mbf{j}\}_{s=1},\{\hat{Q}_\mbf{j}\}_{s=2},\ldots, \{\hat{Q}_\mbf{j}\}_{s=N}\}
\end{equation}
where we have introduced the collective index $\mbf{j}$ corresponding to the length-$N$ Pauli string that describes each operator, i.e. $\hat{Q}_{(021)} = IYX$. The dimension of each weight group is given by
\begin{equation}
    \mathrm{dim}\left(\{\hat{Q}_\mbf{j}\}_{s=k}\right)=\binom{N}{k}3^k.
    \label{eqn:dim_pauli}
\end{equation}

We are interested in situations where the initial operator is a single-site Pauli operator $\hat{O}$, for which \eqref{eqn:Pk_general} takes the form
\begin{equation}
    P_k(t) = \sum\limits_{s(Q_\mbf{j}) = k} \left\lvert f[\hat{Q}_\mbf{j},\hat{O}(t)] \right\rvert^2.
    \label{eq:Pk_pauli}
\end{equation}
Using Eqs.~(\ref{eqn:Pk_haar}) and (\ref{eqn:dim_pauli}) we can evaluate the mean, variance and IPR of this probability distribution for the case of Haar-random evolution. Full expressions are shown in Appendix \ref{appendix:haar}, and their asymptotic behavior is shown in Table \ref{tab:Pk_properties}. 

\begin{table}[t!]
    \centering
    \begin{tabular}{c|c|c}
        & Many spin-$1/2$ & Collective spins \\
        \hline 
        Mean $\mu$ & $\frac{3}{4}N$  & $\frac{2}{3}N$\\
        Variance $\sigma^2$ & $\frac{3}{16}N$ & $\frac{1}{18}N^2$ \\
        IPR $\eta_{\mathrm{IPR}}$ & 
        $\sim N^{-1/2}$ & $\frac{3}{8} N^{-1}$
    \end{tabular}
    \caption{Asymptotic behavior of properties of the operator distribution $\{P_k\}$. Detailed expressions are found in Appendix~\ref{appendix:haar}.}
    \label{tab:Pk_properties}
\end{table}

\subsection{Collective spin systems or singe large spins}
\label{ssec:operator_evolution_collective}
A special case of interest in systems of $N$ spin-$\frac{1}{2}$ particles is when the Hamiltonian is written solely in terms of the collective spin operators $\hat{J}_\alpha = \frac{1}{2}\sum_{i=1}^N \hat{\sigma}_i^{\alpha}$, with $\alpha=x,y,z$. This describes a scenario where the particles show homogeneous all-to-all interactions among themselves, and collective couplings to external fields. Such Hamiltonians preserve the total angular momentum $\hat{J}^2=\hat{J}_x^2+\hat{J}_y^2+\hat{J}_z^2$, and states which are fully symmetric under permutation of particles (corresponding to $J=N/2$) remain so throughout the evolution. Many important models related to quantum chaos and dynamical criticality belong to this class, for instance the Lipkin-Meshkov-Glick model \cite{kochmanski2013,Santos2016}, the quantum kicked top \cite{Haake1987}, and the $p$-spin models \cite{Bapst2012,munoz2021}. \\

The Hilbert space associated with evolution in the symmetric manifold has dimension $d=2J+1=N+1$ and can be spanned by the Dicke states $\{\Ket{J,m}\}$, $m=-J,-J+1,\ldots,J$, which are the eigenstates of $\hat{J}_z$. This space is thus formally equivalent to that of a single particle of spin $J$. While products of angular momentum operators can be used to span any operator in this space, a more convenient choice is given by the spherical tensor operators $\hat{T}_{LM}$, which for an arbitrary $J$ have the form \cite{Klimov2002}
\begin{equation}
    \hat{T}_{LM} = \sqrt{\frac{2L+1}{2J+1}} \sum\limits_{m,m'=-J}^{J} C_{Jm',LM}^{Jm} \KetBra{Jm}{Jm'}
\end{equation}
where $C_{Jm',LM}^{Jm} = \BraKet{Jm',LM}{Jm}$ is the Clebsch-Gordan coefficient which couple two representations of spin $J$ (projection $m'$) and $L$ (projection $M$) to a total spin $J$. The usual selection rules indicate that $m=M+m'$, and so the sum above is restricted to $m-m'=M$. The indices of $\hat{T}_{LM}$ are typically referred to as the rank $L$, which is such that $L-J\leq J \leq L+J$ and hence $0\leq L \leq 2J$, and the projection $M=-L,-L+1,\ldots,L$. 
Spherical tensor operators form the basis for the spin coherent state Wigner function, a generalization of the Wigner function for the harmonic oscillator \cite{agarwal1981relation}.

The spherical tensor operators form an orthonormal operator basis $\trace{T_{L_1 M_1}^\dagger T_{L_2 M_2}}=\delta_{L_1,L_2}\delta_{M_1,M_2}$, and they are in general non-Hermitian with the property $T_{L,M}^\dagger = (-1)^M T_{L,-M}$. The low-rank elements are readily associated with familiar operators
\begin{equation}
    \hat{T}_{1,1} = \alpha_{1,1} \hat{J}_+;\ \hat{T}_{1,0} = \alpha_{1,0} \hat{J}_z;\ \hat{T}_{1,-1}=\alpha_{1,-1} \hat{J}_-,
\end{equation}
where we have omitted the positive normalization constants $\alpha_{LM}$ to lighten the notation. Higher-rank elements correspond to higher-order products of collective spin operators, and can be constructed (see for instance Appendix C of \cite{Chinni2022}) by noting that $\hat{T}_{L,L}=(-1)^L\alpha_{L,L}\hat{J}_+^L$ and using the commutation relations \cite{Sakurai1995} \correc{(we set $\hbar=1$ throughout the paper)} 
\begin{align}
    [\hat{J}_z,\hat{T}_{LM}] & = M\:  \hat{T}_{LM} \\
    [\hat{J}_{\pm},\hat{T}_{LM}] & = \sqrt{(L\mp M)(L\pm M+1)} \hat{T}_{L,M\pm 1}
\end{align}

Physical Hamiltonians are typically written only in terms of low-rank operators (such that $L\ll N$, say), a fact that applies both to actual many-body collective systems and single multi-level atoms \footnote{For example, a natural Hamiltonian for multilevel atoms consists of rotating magnetic fields and a tensor light shift that can be written as sum of spherical tensors up to rank $L=2$ \cite{deutsch2010quantum}}
\cite{deutsch2010quantum}. Following the discussion of the previous section, we will take the rank as the index defining a notion of complexity of any basis operator, where $\mathrm{rank}(\hat{T}_{LM}) = L$. This leads to a grouping of the basis set as 
\begin{equation}
    \{\hat{T}_{LM}\} \rightarrow \{ \{\hat{T}_{L=1,M}\},\{\hat{T}_{L=2,M}\},\{\hat{T}_{L=3,M}\},\ldots\}.
\end{equation}

We will consider our initial operator to be rank-1, e.g. $\hat{O}(0) =\hat{J}_z$, and so the probability distribution in Eq. (\ref{eqn:basis_group}) takes the form
\begin{equation}
    P_k(t) = \frac{1}{\trace{\hat{J}_z^2}} \sum\limits_{M=-k}^k \left\lvert f[\hat{T}_{L=k,M},\hat{J}_z(t)] \right\rvert^2.
    \label{eq:Pk_collective}
\end{equation}
As before, we can use that the dimension of each subset $C_k=\{T_{L=k,M}\}$ is $\mathrm{dim}(C_k)=2k+1$ to compute the properties of $P_k$ for the case where the evolution is Haar-random. Full expressions are given in Appendix \ref{appendix:haar}, and their asymptotic behavior with $N$ is indicated in Table \ref{tab:Pk_properties}. Note that these results admit a direct comparison to the many-body case if one recalls that $J=N/2$. Moreover, $\mathrm{rank}(\hat{T}_{LM})=L$ implies that $\hat{T}_{LM}$ contains up to the $L$th powers of the angular momentum operators $\hat{J}_\alpha = \frac{1}{2}\sum_{i=1}^N \hat{\sigma}_i^{\alpha}$, and is thus composed by up to size-$L$ Pauli operators. The results in Table \ref{tab:Pk_properties} show that, under random evolution, collective spin systems reach smaller operator sizes on average, but lead to broader distributions with variance scaling as $N^2$ instead of $N$.

\section{Scrambling and chaos in the tilted-field Ising model}
\label{sec:scrambling_and_chaos_in_the_tilted_field_Ising_model}

\begin{figure*}[t]
\includegraphics[width=1\textwidth]{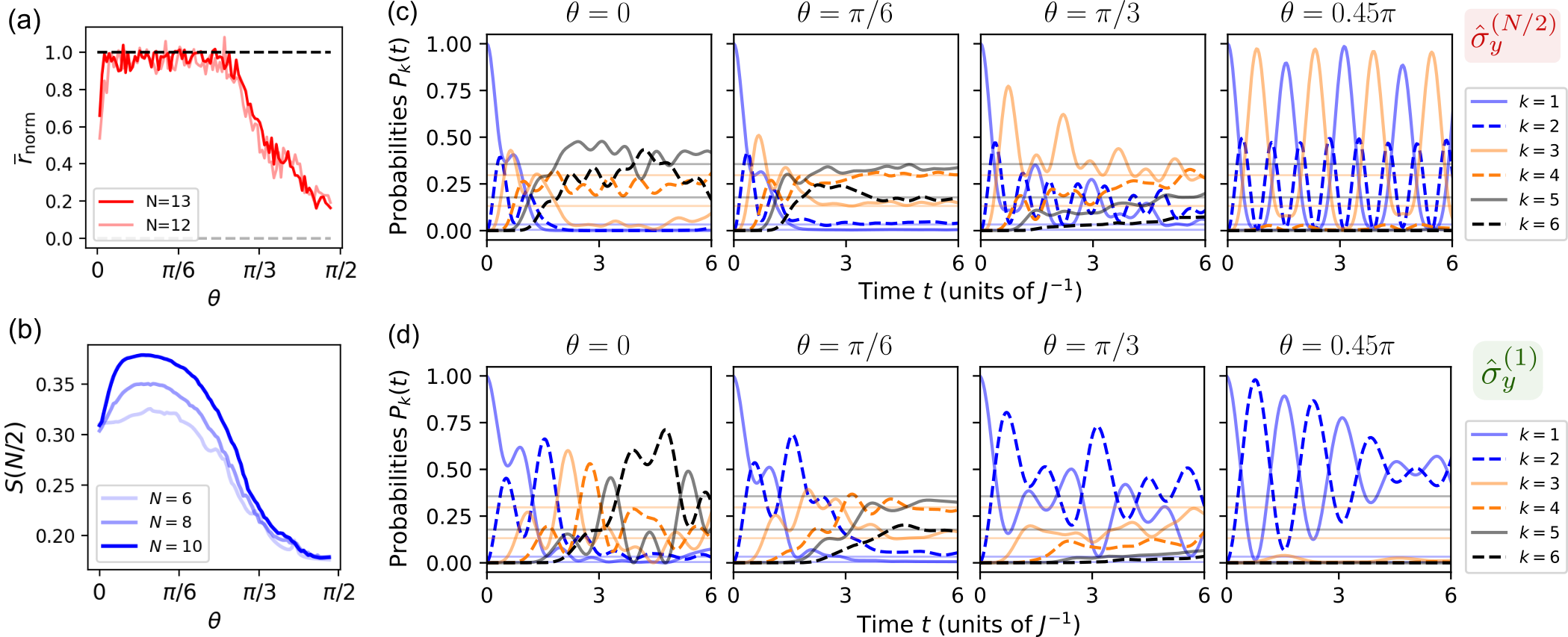}

    \caption{Scrambling from operator evolution and quantum chaos indicators in the tilted field Ising model defined in Eq.~(\ref{eq:tilted field Ising model}), as a function of the external field angle $\theta$. All cases use $B=J$. (a) Normalized mean adjacent level spacing ratio, a standard measure of quantum chaos discussed in Appendix \ref{appendix:avg_ratio}, computed for two instances of the Ising Hamiltonian with different system size $N$. (b) Alternative proxy for quantum chaos given by the averaged entanglement entropy of the eigenstates of $H_{\mathrm{Ising}}$ located in the bulk of the spectrum. (c) and (d) \correc{Short-time evolution} of the coarse-grained operator distribution $P_k(t)$, defined in Eq. (\ref{eq:Pk_pauli}), for different values of $\theta$ and with $N=6$. Cases shown correspond to (c) $\hat{O}(0)=\hat{\sigma}_y^{(N/2)}/\sqrt{2}$ and (d) $\hat{O}(0)=\hat{\sigma}_y^{(1)}/\sqrt{2}$. \correc{thin faint} lines indicate the values of the Haar-random evolution, given in Eq. (\ref{eqn:Pk_haar}) and Eq. (\ref{eqn:dim_pauli}). From left to right, cases corresponding to different values of $\theta \in [0,\pi/2)$ are shown.}
    \label{fig:figure2}
\end{figure*}

We begin by studying the properties of the operator distribution $\{P_k(t)\}$ in the different regimes of the Ising model, a standard paradigm in the study of many-body quantum systems \cite{edwards2010,bernien2017probing,zhang2017observation}. This model describes a set of $N$ spin-$\frac{1}{2}$ particles interacting in $1$D via nearest-neighbor interactions and in the presence of an external magnetic field with a transverse and a longitudinal component. The Hamiltonian can be written as 
\begin{equation}
    H_{\mathrm{Ising}}(\theta)=J\sum_{n=1}^{N-1}\sigma_n^z\sigma_{n+1}^z+B\sum_{n=1}^{N} (\sigma_n^x \cos \theta +\sigma_n^z \sin \theta) ,
    \label{eq:tilted field Ising model}
\end{equation}
where we take $0\leq \theta \leq \frac{\pi}{2}$. Here $\sigma_n^{\alpha}$ are the usual Pauli operators on site $n$ with $\alpha=x,y,z$. For $\theta=0$ \eqref{eq:tilted field Ising model} is the transverse-field Ising model (TIM), whose equilibrium and nonequilibrium properties have been studied extensively \cite{sachdev1999quantum}.
This model is integrable since it can be mapped to a noninteracting system of fermions via the Jordan-Wigner transformation \cite{jordan1993paulische,sachdev1999quantum}. The case of pure longitudinal field $\theta=\frac{\pi}{2}$ is diagonal in the computational basis and thus also trivially integrable. For other values of $\theta$ (and generic choices of $B/J$), this ``tilted-field'' Ising model is quantum chaotic as revealed by several of the usual metrics, which have been studied in previous works \cite{prosen2002general,PhysRevA.71.062324,karthik2007entanglement}. For completeness, we revisit some of those results here. First, the eigenenergies of $H_{\mathrm{Ising}}$ in this chaotic regime display level repulsion as predicted by random matrix theory, and this can be quantified by the average adjacent spacing ratio $\overline{r}$, the details of which we present in Appendix~\ref{appendix:avg_ratio}. We plot a normalized version of this quantity, $\overline{r}_{\mathrm{norm}}$, as a function of $\theta$ in Fig. \ref{fig:figure2} (a). Values close to $1$ indicate agreement with the Gaussian Orthogonal Ensemble (GOE) predictions and thus quantum chaotic behavior, while deviations towards $0$ indicate uncorrelated level statistics typical of integrable systems. As an additional metric, we also consider the average entanglement entropy of the excited states of $H_{\mathrm{Ising}}$ for an equal bipartition of the chain \cite{karthik2007entanglement}. This is defined as
\begin{equation}
    \overline{S}\left(\frac{N}{2}\right) = \sum\limits_{\{\Ket{\phi_i}\}} S\left(\rho^{(i)}_{\frac{N}{2}}\right)
\end{equation}
where $\rho^{(i)}_{\frac{N}{2}}$ is the state resulting from tracing out half of the particles from $\KetBra{\phi_i}{\phi_i}$, $\Ket{\phi_i}$ is an eigenstate of $H_{\mathrm{Ising}}$, and the sum is carried out over the bulk the spectrum (i.e. avoiding the ground and low energy states of $H_{\mathrm{Ising}}$ and $-H_{\mathrm{Ising}}$). Regimes of maximum average bipartite entanglement are then associated with quantum chaos, as can be verified from the results shown in Fig. \ref{fig:figure2} (b).

We now study the scrambling process in the different regimes of the Ising model by analyzing the probability distribution $P_k(t)$ defined in Eq. (\ref{eq:Pk_pauli}). In Fig.~\ref{fig:figure2} (c) and (d) we display exact numerical results of the time-dependent distribution corresponding to a chain of $N=6$ particles where the initial operator sits either  (c) in middle of the chain $\hat{O}(0)=\hat{\sigma}_y^{(N/2)}/\sqrt{2}$, or (d) at the edge $\hat{O}(0)=\hat{\sigma}_y^{(1)}/\sqrt{2}$. In both cases, the initial distribution is initially concentrated in $P_1(0)=1$ and then evolves in time displaying different features depending on the system parameters. All cases with $0\leq \theta \lesssim \frac{\pi}{4}$ show a rapid decrease of the initial component $P_1$ as the operator spreads into a superposition of larger-size configurations. Crucially, however, the chaotic case $\theta=\pi/6$ shows a fast equilibration to the values corresponding to the random distribution shown as dashed lines, cf. Eq.~(\ref{eqn:Pk_haar}). As $\theta\rightarrow 0$ and the model becomes integrable, the distribution shows further oscillations and fails to equilibrate completely in the timescale shown. The deviations from ergodicity are enhanced when the initial operator sits at the edge of the chain, a situation in which the initial configuration has less options to equilibrate to since the site has a single neighbor instead of two due to the open boundary conditions.

The other integrable regime, occurring at $\theta = \pi/2$ but already noticeable for $\theta=\pi/3$, corresponds to a very different type of evolution. In the diagonal case the external field commutes with the interaction, and thus a single site operator like $\hat{\sigma}_y^{(l)}$ evolves to only two- and three- site operators, independent of the length of the chain. The distribution then shows very little spreading as most of the elements are never populated. The situation remains roughly the same even in the presence of a small transverse field, as can be seen in the cases corresponding to $\theta=0.45\pi$ shown in Fig.~\ref{fig:figure2}.

\begin{figure}
\includegraphics[width=1\linewidth]{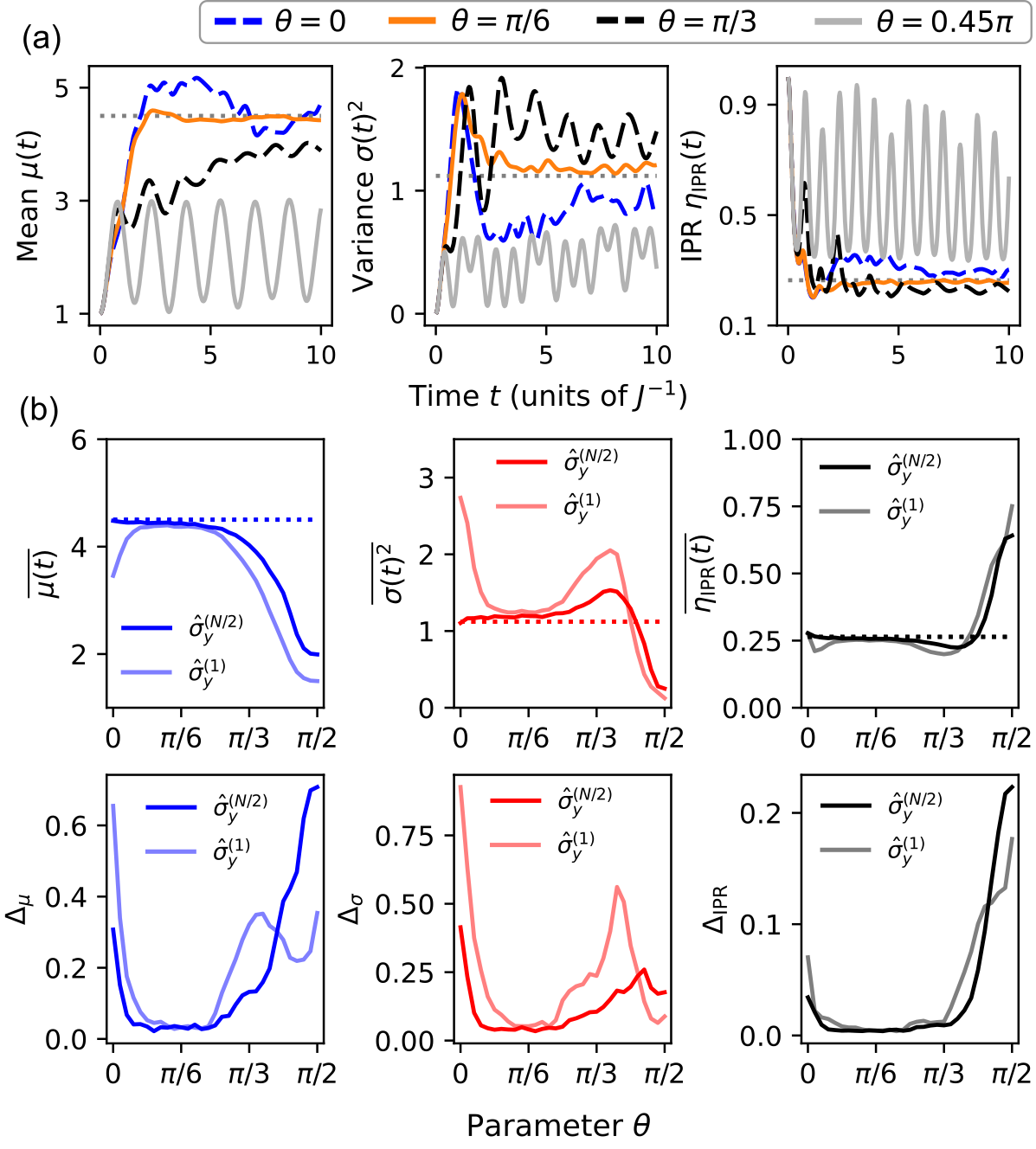}
    \caption{Measures of the probability distribution $\{P_k\}$ for the tilted field Ising model with $N=6$ and $B/J=1$. (a) Mean $\mu(t)$, variance $\sigma^2(t)$, and IPR $\eta_{\mathrm{IPR}}(t)$ of the operator distribution are shown as a function of time for the four values of $\theta$ displayed in Fig.~\ref{fig:figure2}. \correc{Dotted} lines correspond to the predictions for Haar-random evolution shown in Table \ref{tab:Pk_properties}. Results shown here correspond to $\hat{O}(0)=\hat{\sigma}_y^{(N/2)}/\sqrt{2}$.  (b) Long-time properties of the aforementioned measures, as calculated by the time-averaged value, cf. Eq.~(\ref{eq:time_avg}), and the time-averaged temporal fluctuations, cf. Eq.~(\ref{eq:time_fluct}). Plots are shown as a function of the parameter $\theta$ of the Ising model, and for choices of the initial operator on the middle site (dark lines) and at the edge (light lines) of the chain.}
\label{fig:figure3}
\end{figure}

While the operator probability distribution $\{P_k(t)\}$ already reduces the description of observable evolution from the exponentially large basis to a set of only $N$ numbers, it is still helpful to analyze measures which describe particular aspects of the distribution at each time. We thus turn to study the quantities introduced in Sec. \ref{sec:operator_evolution}, namely the mean $\mu(t)$, variance $\sigma(t)$, and IPR $\eta_{\mathrm{IPR}}(t)$ of the distribution. Figure~\ref{fig:figure3}~(a) shows the evolution of these three quantities for different values of $\theta$ for the initial operator located in the middle of the $N=6$ particle chain (for completeness, the case where the initial operator sits at the edge is shown in Appendix~\ref{appendix:additional}). The evolution of the mean $\mu(t)$ shows a increase from $\mu(0)=1$ towards $\sim N$ with different features depending on the value of $\theta$. The most chaotic case, $\theta=\pi/6$, grows until reaching the random value $3N/4$ (see Table \ref{tab:Pk_properties}), while most other cases show oscillations. Note that the TIM case ($\theta=0$) grows beyond the random value, meaning that during certain periods, the integrable model leads to mean operator sizes which are larger than those found for Haar-random evolution. Finally, as the other integrable limit is approached for $\theta\sim \pi/2$, the distribution stops shifting beyond $N=3$, as discussed before, and shows indefinite large-amplitude oscillations. 

For the evolution of the variance of the distribution $\sigma^2(t)$, we observe that cases far from the diagonal case (i.e. $\theta \ll \pi/2$) show an initial increase from $\sigma^2(0)=0$ which shoots up significantly above the Haar-random prediction $\sim 3N/16$. Then, most cases equilibrate to that value, albeit in different timescales. Interestingly, as $\theta$ reaches $\pi/3$ the variances become consistently larger than the Haar prediction. This indicates a nontrivial behavior in which not necessarily the most chaotic evolution leads to the largest width of the operator distribution. A similar trend is observed in the IPR of the distribution, $\eta_{\mathrm{IPR}}(t)$, shown also in Fig.~\ref{fig:figure3} (a). We point out that $\mu(t)$ and $\sigma(t)$ as a function of time for the chaotic case $\theta=\pi/6$ show a remarkable similarity to the ones observed for the SYK model by Roberts \textit{et al.} in Ref.~\cite{Roberts2018}. 

In order to obtain a more general picture of how the different regimes of the Ising model with varying $\theta$ are reflected in the properties of operator-size distribution, we analyze the long-time behavior of these measures $X(t)\in\{\mu(t),\sigma(t),\eta_{\mathrm{IPR}}(t)\}$ by computing both the time-averaged value
\begin{equation}
    \overline{X(t)} \equiv \frac{1}{t_f}\int\limits_{t_0}^{t_f} X(t')dt'
    \label{eq:time_avg}
\end{equation}
\noindent and the time-averaged temporal fluctuations,
\begin{equation}
    \Delta_X^2 \equiv \frac{1}{t_f}\int\limits_{t_0}^{t_f} \left(X(t')-\overline{X(t)}\right)^2 dt'.
    \label{eq:time_fluct}
\end{equation}
For all numerical calculations we integrate the quantities from $t_0\neq0$ such that the initial transient does not contribute, and take $t_f\gg t_0$ to estimate the infinite-time average in each case. In Fig.~\ref{fig:figure3} (b) we show the time-averaged value and time-averaged temporal fluctuations for each of the three measures as a function of $\theta$ and for initial operators in both the middle and at the edge of the chain. The results are for $N=6$, $Jt_0=5$ and $Jt_f=40$, but they do not depend significantly on these choices. There are two overarching features that stand out clearly in all cases shown. First, the quantum chaotic regime spanning roughly the $\theta\in(0,\pi/3)$, as seen in Fig.~ \ref{fig:figure2} (a) and (b), yields i) long-time equilibration of all the distribution measures to the Haar-random predictions, as can be seen from the agreement between the time-averaged values in the first row with with the Haar predictions from Table \ref{tab:Pk_properties}, and ii) suppression of temporal fluctuations $\Delta_X\simeq 0$ for all cases. Second, the `trivially' integrable regime, which is reached as $\theta\rightarrow \pi/2$, can be clearly distinguished from the features of the distribution, since the mean and the spread are greatly reduced as the distribution tends to be confined to $k=1,2,3$ independent of system size. Moreover, this highly nonergodic case leads to greatly enhanced temporal fluctuations, as seen in particular from the behavior of $\Delta_\mu$ and $\Delta_{\mathrm{IPR}}$.

On the other hand, we observe that the studied properties of the distribution have a harder time distinguishing the integrable model at $\theta\rightarrow 0$ from the ergodic case. For instance, the time-averaged values of the mean, variance, and IPR stay very close to the random predictions as $\theta\rightarrow 0$, indicating that the integrable transverse Ising model leads to significant, random-like operator spreading at long times. This is true for most choices of the initial operator; however, we find that choosing $\hat{O}(0)$ at the edge leads to somewhat different features, particularly in the $\theta\simeq 0$ regime. For this regime and choice of initial operator, we see that the time-averaged mean drops and the time-averaged variance rises, signaling a clear deviation from ergodicity. Interestingly, while the behavior of the mean is similar to the other integrable limit, the case of the variance is opposite -- the width of the distribution increases above the chaotic case in the transverse-field regime. 

More generally we observe that the integrability of the model at $\theta=0$ consistently leads to increased long-time temporal fluctuations in all quantities, independent of the choice of the initial operator. Enhanced oscillations in the mean, variance, and IPR are seen in both integrable regimes as compared to the chaotic case. We thus find that these temporal fluctuations can be used to distinguish chaotic and nonchaotic regimes of the model. These results are aligned with the findings of Ref.~\cite{fortes2019}, who showed that the long-time behavior (particularly the properties of the frequency spectra) of OTOCs serves as a good indicator to distinguish quantum chaos for integrability in various models, including the Ising model considered here. In Sec.~\ref{sec:otocs_and_connections} we will further discuss the connection between the quantities studied here and OTOCs.  

Finally, we point out that the results shown here for a fixed system size of $N=6$ are representative of other cases, which we show in Appendix~\ref{appendix:additional}. In particular we show in Fig. \ref{fig:figure_app_N} that cases with $N=5,6,7$ behave very similarly and essentially coincide (when properly normalized) in the chaotic regime. We also observe that the magnitude of temporal fluctuations decay with increasing system size, a behavior typical of ergodic systems.

\section{Scrambling and chaos in the Quantum Kicked Top}
\label{sec:scrambling_and_chaos_in_the_quantum_kicked_top}

We now turn our attention to the study of operator spreading and scrambling in collective spin systems, which we introduced in Sec.~\ref{ssec:operator_evolution_collective}. We will consider the dynamics of a quantum kicked top (QKT), a paradigmatic model of quantum chaos first introduced by Haake, Ku\'s, and Scharf \cite{Haake1987}, which has been the subject of many theoretical  \cite{Schack1994, Sieberer2019, Ghose2008} and experimental \cite{Chaudhury2009,neill2016,Lysne2020} studies. The QKT time evolution operator of interest for this study can be written as
\begin{equation}
    \hat{U}_{\mathrm{QKT}} = \hat{U}_z \hat{U}_y \hat{U}_x,\ \mathrm{with}\     \hat{U}_{\mu}=e^{-i\left(\alpha_\mu \hat{J}_\mu + \frac{\gamma_\mu}{2J} \hat{J}_\mu^2\right)},
    \label{eq:qkt_unitary}
\end{equation}
where $\mu=x,y,z$ and the total angular momentum is $J=2N$. The model in Eq.~(\ref{eq:qkt_unitary}) is constructed in such a way as to avoid parity and time-reversal symmetry, which are present in the original QKT of Ref.~\cite{Haake1987}; see also \cite{Kus1988} for the study of similar models. Each of the unitaries $\hat{U}_\mu$ can be regarded as generated by a ``twisting and turning'' collective Hamiltonian \cite{Sieberer2019}, composed of a rotation term $\hat{J}_\mu$ and a twisting or interaction term $\hat{J}^2_\mu=\sum_{ij}\hat{\sigma}_i^{(\mu)}\hat{\sigma}_i^{(\mu)}/4$. Since the symmetric subspace dimension scales linearly with the number of particles $d=N+1$, large values of $N$ can be accessed numerically in these types of systems. 

\begin{figure}
    \centering
    \includegraphics[width=0.95\linewidth ]{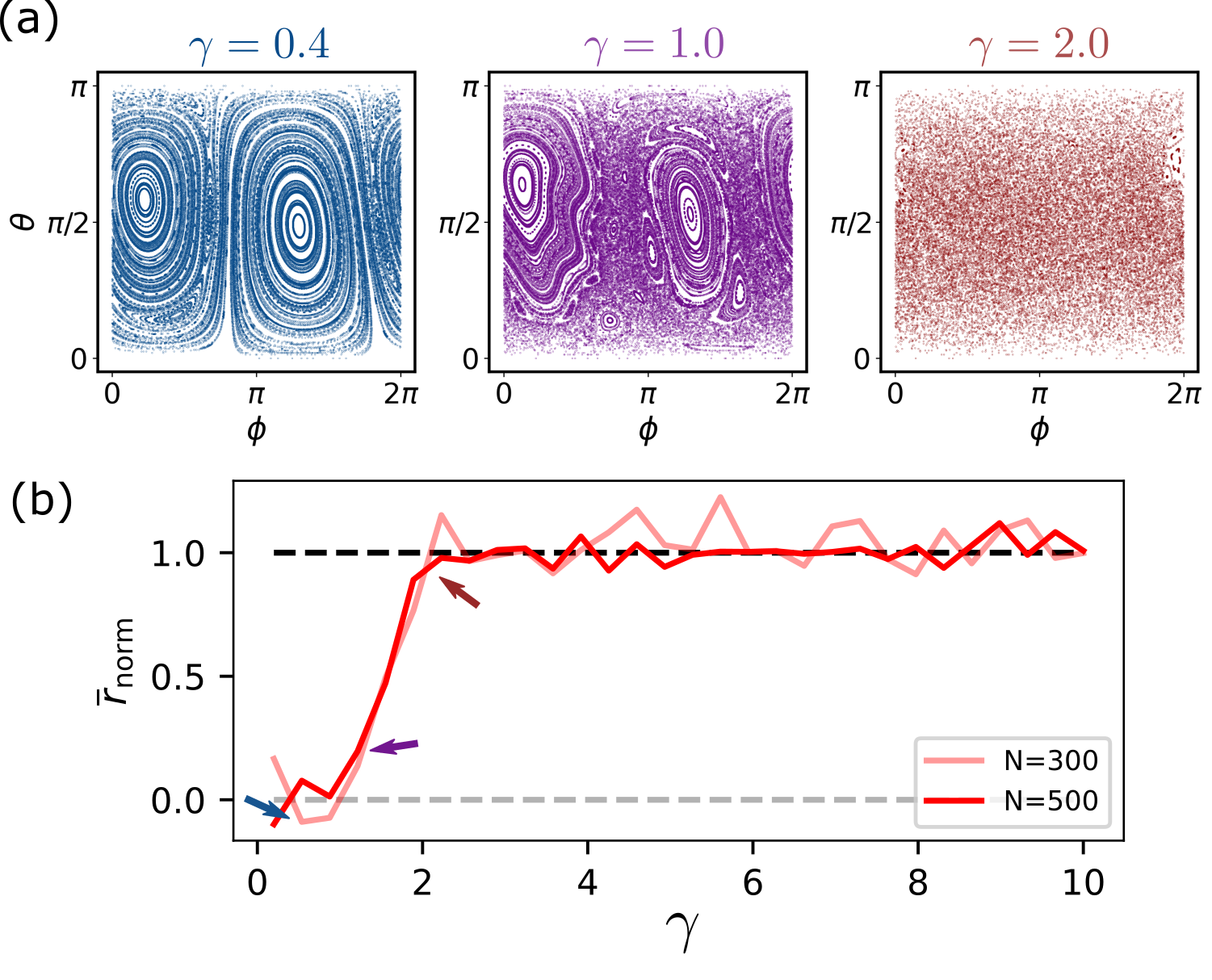}
    \caption{
    Quantum and classical chaos in the kicked top model of Eq.~(\ref{eq:qkt_unitary}), with parameters $\alpha_x=1.7$, $\alpha_y=1$, $\alpha_z=0.8$ and $\gamma_x=0.85\gamma$, $\gamma_y=0.9\gamma$ and $\gamma_z=\gamma$. (a) Phase space portraits (Poincaré sections) of the classical kicked top for different values of the nonlinearity parameter $\gamma$. The system transitions from a regular regime at small $\gamma$, to a mixed phase space for $\gamma \sim 1$ to a full chaotic regime for $\gamma \gtrsim 2$. (b) Normalized mean adjacent level spacing ratio for two instances of the QKT with different $N=2J$. Note that due to the lack of time-reversal symmetry of this model, the measure is normalized to give 1 when the Circular Unitary Ensemble (CUE) value is achieved. See Appendix \ref{appendix:avg_ratio} for more details.}
    \label{fig:figure_qkt_1}
\end{figure}

The quantum chaotic properties of the QKT can be fully understood by studying the associated classical kicked top, which can be recovered as the mean-field limit of the map generated by Eq.~(\ref{eq:qkt_unitary}) \cite{Haake1987}. The resulting classical area-preserving map acts on a spherical phase space whose coordinates are $\mbf{R}\equiv (X,Y,Z)=\lim_{J \to \infty}\langle \hat{\mbf{J}}\rangle /J$. In Fig.~\ref{fig:figure_qkt_1} (a) we show the Poincaré sections corresponding to this map, where we have chosen the system parameters to be $\alpha_x=1.7$, $\alpha_y=1$, $\alpha_z=0.8$ and $\gamma_x=0.85\gamma$, $\gamma_y=0.9\gamma$ and $\gamma_z=\gamma$. For $\gamma=0$ the system is trivially integrable as $\hat{U}_{\mathrm{QKT}}$ generates only rotations, and as $\gamma$ is increased the classical phase space becomes mixed, with islands of regular motion separated by areas of chaos. For $\gamma\gtrsim 2$, most of the phase space becomes chaotic. This transition to chaos can be clearly observed from the normalized average adjacent spacing ratio $\overline{r}_{\mathrm{norm}}$, introduced in the previous section (see also Appendix \ref{appendix:avg_ratio}), which we show in Fig.~\ref{fig:figure_qkt_1} (b). 

\begin{figure}[t!]
    \centering
    \includegraphics[width=\linewidth ]{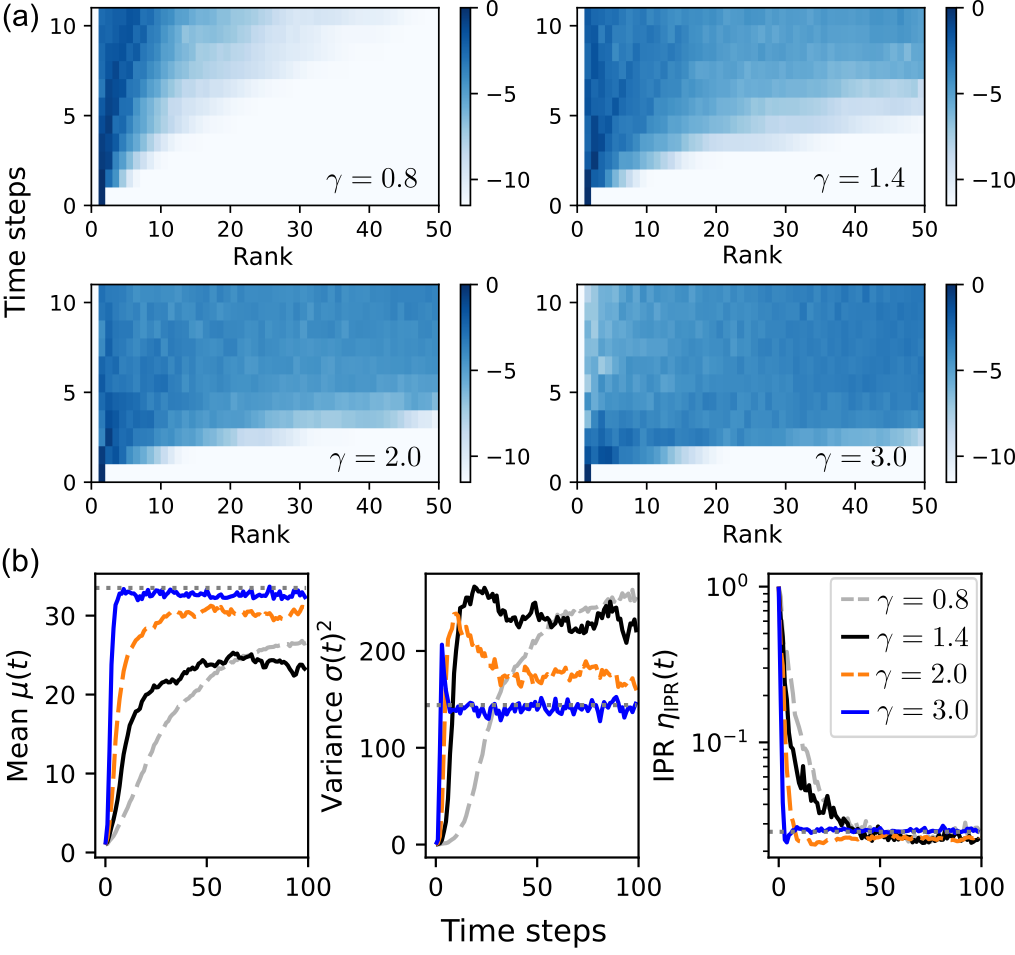}
    \caption{Dynamics of the operator distribution for the QKT of Eq.~(\ref{eq:qkt_unitary}). Results were obtained using parameter values for $\{\alpha_i,\gamma_i\}$ identical to Fig.~\ref{fig:figure_qkt_1}, $N=2J=49$, and the initial operator $\hat{O}(O)=\hat{J}_z$. (a) Density plots of the operator distribution $P_k(t)$, shown in logarithmic scale for clarity -- the quantity plotted is $\log \left(P_k(t)+\epsilon \right)$, with $\epsilon=10^{-10}$. (b) Measures of the distributions shown in (a), calculated as a function of time.}
    \label{fig:figure_qkt_2}
\end{figure}

\begin{figure}[t!]
    \centering
    \includegraphics[width=\linewidth ]{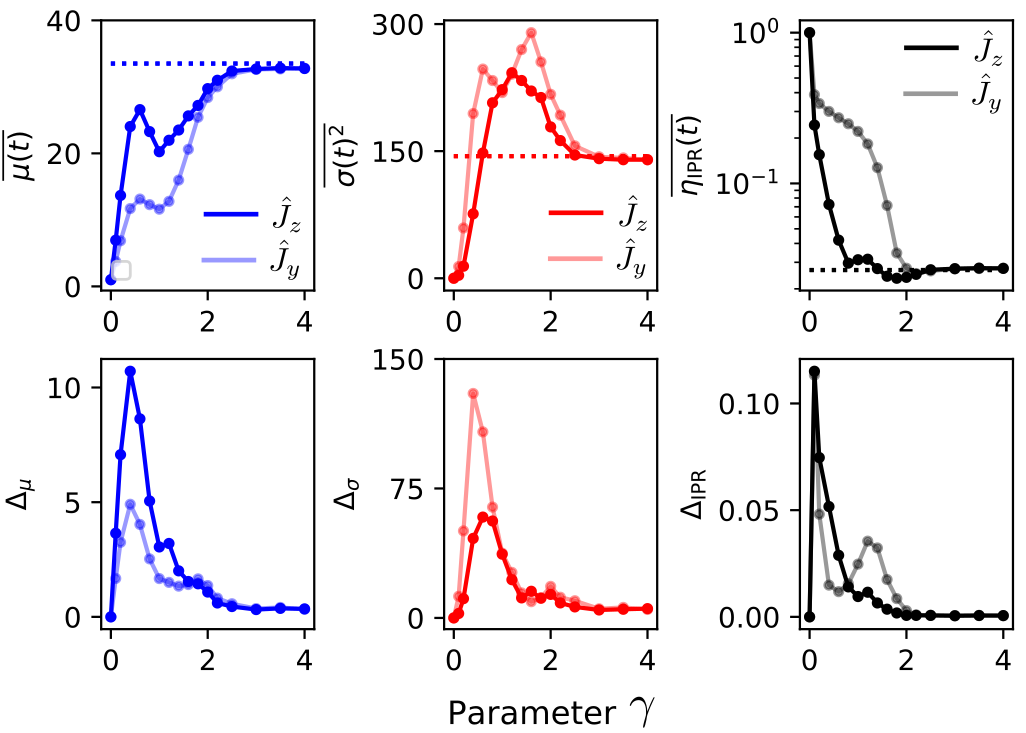}
    \caption{Long-time averages (top row), cf. Eq.~(\ref{eq:time_avg}) and averaged temporal flucutations (bottom row), cf. Eq.~(\ref{eq:time_fluct}), for the mean, variance, and IPR of the operator distribution for the QKT. Results are shown for $N=2J=49$ using identical parameter values for $\{\alpha_i,\gamma_i\}$ as Fig.~\ref{fig:figure_qkt_1}, and for two distinct initial operators $\hat{O}(0)=\hat{J}_z$ (dark lines in all plots), and $\hat{O}(0)=\hat{J}_y$ (light lines). All dashed lines correspond to the predictions for Haar-random evolution shown in Table~\ref{tab:Pk_properties}.}
    \label{fig:figure_qkt_3}
\end{figure}

Being able to access large system sizes also means that, even when coarse grained, it can be hard to visualize the evolution of each component of the probability distribution $\{P_k(t)\}$ defined in Eq.~(\ref{eq:Pk_collective}) in a manner similiar to what was done for a small instance of the Ising model in Fig.~\ref{fig:figure2}. In Fig.~\ref{fig:figure_qkt_2} (a) we present density plots of the distribution at short times for four representative values of $\gamma$ and an initial choice of operator $\hat{O}(0)=\hat{J}_z$. Each horizontal slice corresponds to a snapshot of the distribution at a given time, while vertical slices show the evolution of individual components. The plots illustrate how the distribution, which is concentrated at rank $k=1$ at $t=0$, spreads onto higher ranks faster as $\gamma$ is increased. In order to capture the long-time properties of this evolution, we display in Fig.~\ref{fig:figure_qkt_2} (b) the mean $\mu(t)$, variance $\sigma(t)$, and IPR $\eta_{\mathrm{IPR}}(t)$ of the operator distribution as a function of time. As expected, we observe how the values predicted for Haar-random evolution (dashed lines) are readily attained as $\gamma$ increases and the QKT becomes chaotic. We also observe an interesting similarity between the behavior of both the mean and the variance when compared to the Ising case, cf. Fig.~\ref{fig:figure_qkt_3} (a). In the ergodic cases the mean $\mu(t)$ increases steadily and saturates at the prediction from Table~\ref{tab:Pk_properties}, but the variance temporarily `shoots up' before it equilibrates. We emphasize that this same behavior has been observed for the SYK model in a previous work \cite{Roberts2018}. The fact that the QKT, which is essentially a quantized classical system, also displays similar features is indicative of the presence of unifying features in the evolution of the operator distribution, even when considering quantum chaotic models of very different natures. Interestingly, we observe in both the QKT and the Ising model that the variance of the distribution can be systematically larger in the non-chaotic regime with respect to the chaotic case.

Finally, we study the long-time averages Eq.~(\ref{eq:time_avg}) and time-fluctuations Eq.~(\ref{eq:time_fluct}) of the different measures of the operator probability distribution as a function of the $\gamma$ parameter in the QKT, analogous to the results presented for the Ising model in Fig.~\ref{fig:figure3} (b). Results for the QKT are shown in Fig.~\ref{fig:figure_qkt_3} for two choices of operators: $\hat{O}(0)=\hat{J}_z$ (dark lines) and $\hat{O}(0)=\hat{J}_y$ (light lines). Similar to the results obtained for the Ising model, in the chaotic regime of the QKT the time-averaged mean, variance, and IPR closely match the predictions from Haar-random evolution, with vanishing temporal fluctuations. In the opposite (integrable) regime, the trivial dynamics at $\gamma\sim0$ shows mostly localized distributions and little spreading (and, correspondingly, small fluctuations), akin to the regime of $\theta\sim \pi/2$ of the Ising model.

In the transition to chaos, for $0\lesssim \gamma \lesssim 2$, the properties of the distribution show some unifying features, but is overall operator-dependent. We observe that the initial operator $\hat{O}(0)=\hat{J}_z$ shows significantly more spreading than $\hat{J}_y$. Analyzing the classical phase spaces in Fig.~\ref{fig:figure_qkt_1} (a), one readily observes that stable islands tend to be localized on the equator of the spherical phase space, with unstable areas around the poles. Around these unstable fixed points is where chaos first appears already at $\gamma=1.0$ \cite{reichl1992,chinni2021}. We attribute the enhanced growth of $\hat{J}_z$ to these instabilities, in a phenomenon closely related to the previously studied saddle-point scrambling \cite{kidd2021saddle}. Aside from this we find for both operators that the transition to chaos is characterized by an enhanced variance of the distribution. Interestingly, the shape of $\overline{\sigma(t)^2}$ for the QKT in the regime $0\leq \gamma \leq 2$ is very similar to that of the Ising model in the equivalent regime $\pi/4\lesssim \theta \leq \pi/2$. This similarity once again hints at a unified behavior of the operator growth in the transition from nonergodic to ergodic behavior. 

Finally, we find that temporal fluctuations (bottom row of Fig.~\ref{fig:figure_qkt_3}) in the QKT are a good proxy for quantum chaos in the model, similar to what we observed earlier for the Ising model. However, in this case the correspondence does not extend to very small values of $\gamma$, since the trivial dynamics of the QKT leads to a roughly constant operator distribution. As in the Ising model we find for the chaotic case of the QKT that the temporal fluctuations decrease with system size while the time-averaged measures are roughly independent of the values of $N$ (see Appendix~\ref{appendix:additional}).

\section{Scrambling in a quantum circuit model}
\label{sec:Scrambling in circuit model}

\begin{figure*}[t!]

\includegraphics[width=1.\textwidth]{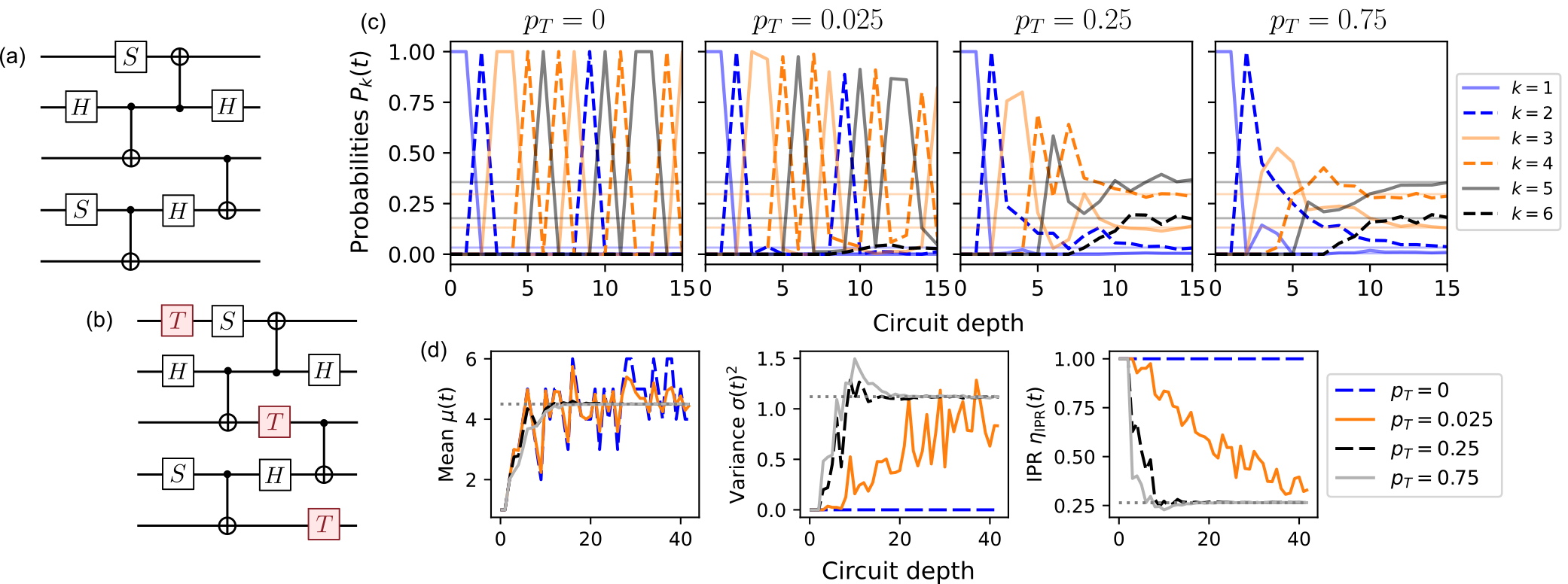}
\caption{Scrambling and operator growth in a random quantum circuit. (a) One instance of a random circuit composed only of the Clifford gates $H$, $S$ and $CX$ ($p_T=0$). (b) Same instance with non-Clifford $T$ gates interleaved in random locations. (c) Evolution of the operator distribution probabilities $P_k(t)$ for different values of the $T$-gate probability $p_T$. From the left to right the ``non-Cliffordness'' of the circuit increases with $p_T$. Results shown for $p_T>0$ are averages over 40 random instances. (d) Evolution of the mean $\mu(t)$, variance $\sigma^2(t)$, and IPR of the operator distribution $\eta_{\mathrm{IPR}}(t)$ as a function of time, and for the same values of $p_T$ as used in (c). The initial operator is $\hat{O}(0)=\sigma_z^{(1)}/\sqrt{2}$.}
    \label{fig:figure_clifford}
\end{figure*}

As a final case study, we analyze how the operator distribution $\{P_k(t)\}$ behaves for a quantum circuit akin to the ones studied in quantum computing \cite{nielsen_chuang_2010,brylinski2002universal}. We consider circuits formed by gates in the universal set $\{H,T,S,CX\}$, where
\begin{equation}
\begin{aligned}
     &H=\frac{1}{\sqrt{2}}\begin{pmatrix}
    1 & 1\\
    1& -1
    \end{pmatrix},\hspace{0.1cm} S=\begin{pmatrix}
    1 &0 \\
    0& i
    \end{pmatrix},\\
    &\hspace{0.1cm} T=\begin{pmatrix}
    1 &0 \\
    0& \sqrt{i}
    \end{pmatrix}, \hspace{0.1cm} 
    CX= \begin{pmatrix}
    1&0&0&0\\
    0&1&0&0\\
    0&0&0&1\\
    0&0&1&0
    \end{pmatrix}
\end{aligned}.
\end{equation}
It is known that combinations of $\{H,S,CX\}$ create a Clifford circuit \cite{gottesman1998heisenberg}, the action of which takes a Pauli operator to another Pauli operator. However, the $T$ gate is not a Clifford gate and thus transforms a Pauli operator to a sum of Pauli operators. 
From the Gottesman-Knill theorem, the evolution of a Clifford circuit can be simulated efficiently on a classical computer \cite{gottesman1998heisenberg}, whereas the presence of $T$ gates make the simulation of the non-Clifford circuit inefficient  \cite{gottesman1998heisenberg,aaronson2004improved,bravyi2016}. Universal quantum computing requires non-Clifford operations such as the $T$ gates, and the role of these operations in information scrambling has been studied in recent works \cite{leone2021}.

Here we focus on how the probability distribution $P_k(t)$ changes as one considers Clifford versus non-Clifford circuits. We study a random circuit model where each layer has a random arrangement of Clifford gates as in Fig.~\ref{fig:figure_clifford} (a), and where, with probability $p_T$, a $T$ gate is applied in the layer. The numerical results are obtained after randomly sampling 40 instances of the circuit for an initial operator $\sigma_z^{(1)}$ for $N=6$ qubits. 
In Fig.~\ref{fig:figure_clifford} (c) each probability $P_k(t)$ is plotted as a function of the circuit depth for different values of $p_T$. 
From our analysis it is clear that if the probability $p_T$ is not too small, the operator distribution becomes the one predicted from the Haar-random evolution after some depth of the circuit, and the required depth gets lower as we increase the $p_T$. This behavior is in line with the fact that including the non-Clifford gates makes the gate set universal, and thus random circuit instances are able to explore uniformly the space of unitaries. On the other hand it is interesting to analyze the behavior at small $p_T$, where the system behaves (roughly) as a Clifford circuit. As can be seen from the evolution of the probabilities $P_k(t)$, this does not preclude the spreading of operators into higher weights. However, since Pauli operators are approximately mapped into Pauli operators, the distribution is very localized at all times. This is the signature of \textit{quasi}-scrambling (as opposed to \textit{genuine} scrambling \cite{zhuang2019scrambling}), also termed operator spreading (as opposed to operator entanglement) \cite{mi2021}.

Finally, in Fig.~\ref{fig:figure_clifford} (d), we analyze the mean, variance, and IPR of the operator distribution for this random circuit model, similarly to the analysis of tilted field Ising model and QKT in the previous sections. For sufficiently large $T$ gate probabilities $p_T$, yielding a large number of $T$ gates, the behavior of these are identical in nature to the one found for other models in the chaotic regime, see Fig.~\ref{fig:figure_qkt_2} (b) and tilted field Ising model in Fig.~\ref{fig:figure3} (a): a sharp increase of the mean and variance and quick equilibration to the random predictions, including the shoot-up of the variance at intermediate times before also equilibrating. For small $p_T$, the quasi-scrambling behavior is clearly seen in these measures: while the mean of the distribution increases in a manner very similar to the other \correc{cases} (albeit with enhanced temporal fluctuations), the variance and IPR show very different behaviors as the distributions localize when $p_T\rightarrow 0$. \correc{This analysis shows how the breaking down of classical tractability at $p_T\neq 0$ can be witnessed, dynamically, by the early-time growth of the operator distribution variance (or, conversely, the decay of its IPR)}.

\section{Connection to OTOCS}
\label{sec:otocs_and_connections}

In Sec.~\ref{sec:operator_evolution} we argued that the properties of scrambling, understood as delocalization of quantum information along the degrees of freedom of a system, are encoded in the coarse-grained operator distribution $\{P_k(t)\}$ defined over a particular partitioning of the operator basis, cf. Eqs.~(\ref{eqn:basis_group}) and (\ref{eqn:Pk_general}). Studies of scrambling in the literature, however, are often focused on the analysis of out-of-time-ordered correlators (OTOCs) of the form
\begin{equation}
    \mathcal{C}\left(\hat{W}(t),\hat{V}\right) = \frac{1}{d}\trace{\hat{W}^\dagger(t)\, \hat{V}^\dagger(0)\, \hat{W}(t)\,\hat{V}(0)},
    \label{eq:otoc_vw}
\end{equation}
where here we consider the OTOC to be evaluated for a thermal state at infinite temperature. In this section we discuss the mathematical connection between the operator distribution and the OTOC by summarizing some previous results in the literature \cite{Roberts2018,qi2019} and showing novel relations.

We focus our attention on the case of systems of spin-$1/2$ particles for simplicity (see Ref.~\cite{zhuang2019scrambling} for a detailed study of systems of qudits using a generalized Pauli basis, and Ref.~\cite{yin2021} for the case of collective systems and kicked tops). We take the operators $\hat{W}$ and $\hat{V}$ in Eq.~(\ref{eq:otoc_vw}) to be in the $N$-qubit Pauli set $\mathcal{P}$, and for our purposes it suffices to think of $\hat{W}(0)$ as a single site operator, i.e. $\hat{W}(0)\in C_1$ using the notation used in Sec.~\ref{ssec:operator_evolution_pauli}. To discuss the connection between this family of OTOCs and the operator size distribution $\{P_k\}$, we define the $n$th moment of the latter as
\begin{equation}
    \mu_n(t) = \sum\limits_{k=1}^N k^n P_k(t).
\end{equation}
\noindent where $\mu_1\equiv \mu$ to be consistent with our choice of notation in previous sections. We then consider the average of OTOCs over the subspace $C_n$ of $n$-body Pauli operators
\begin{equation}
        \mathcal{M}_n(t) = \frac{1}{\dim(C_n)} \sum\limits_{\hat{R}\in C_n} \mathcal{C}(\hat{W}(t),\hat{R}).
        \label{eq:avg_otocs}
\end{equation}
The simplest connection between this quantities is that 
\begin{equation}
    \mu_1(t) = \frac{3N}{4}\left(1-\mathcal{M}_1(t)\;\right)
    \label{eq:connect_otoc1}
\end{equation}
\noindent which has been studied in many previous works \cite{Roberts2018,qi2019,hosur2016}  (for completeness we provide a proof of this relation in Appendix \ref{appendix:otoc}). We point out that a closely related connection can be drawn between the OTOCs and the average cluster size via the spectrum of multiple quantum coherences in an NMR setting \cite{alvarez2010,garttner2017,dominguez2021}. Equation~(\ref{eq:avg_otocs}) shows that the mean operator size can be obtained by measuring $\dim(C_1)=3N$ OTOCs, one per each single site operator $\hat{R}\in C_1$. A perhaps less known relation is that
\begin{equation}
    \mu_2(t) = \frac{9}{16}N(N-1)\left(\mathcal{M}_2(t)-1\right) + \frac{3N-1}{2}\mu_1(t),
    \label{eq:connect_otoc2}
\end{equation}
indicating that in order to determine the variance of the operator distribution $\sigma^2(t)=\mu_2(t)-\mu_1(t)^2$ one now requires additional access to $\sim N^2$ OTOCs on two-body operators $\hat{R}\in C_2$. In Appendix~\ref{appendix:otoc} we show the following general relation
\begin{equation}
    \mathcal{M}_n(t) = \sum\limits_{i=0}^n \alpha_n^{(i)} \mu_i(t) + \alpha_n^{(0)} \label{eq:OTOCConnectionMn}
\end{equation}
which highlights that, in general, reconstructing the $n$th moment of the operator distribution requires us to access averages of OTOCs involving up to $n$ bodies. Moreover, the connection is not straightforward, as the coefficients $\alpha_n^{(i)}$ for larger $i$ take exponentially small values as $n$ and $N$ increase. This implies that reconstructing the \textit{complete} operator probability distribution $\{P_k\}$ via OTOCs is an unfeasible task, even if one were able to easily access OTOCs experimentally.

However, one might argue that for many situations of interest, merely obtaining lower order moments like $\mu_1$ and $\mu_2$ would be sufficient. In an experimental setup, one is then confronted with the fact that OTOCs are intrinsically hard to access and typically require auxiliary systems \cite{landsman2019}, time-reversal operations \cite{garttner2017,PhysRevX.7.031011,swingle2016}, or statistical correlations through randomized measurements \cite{vermersch2019,PhysRevLett.124.240505}, and clearly measuring $\sim N^2$ hard objects is undesirable. Notice that one might also invoke arguments of self-averaging, i.e., measuring a single choice of $\hat{R}$ in Eq.~(\ref{eq:connect_otoc1}) might give a satisfactory indicator of the behavior of the averaged OTOC (and thus of the mean operator size) if the system is sufficiently ergodic. However, as we have shown in this work, the properties of the operator distribution are able to distinguish interesting features of the behavior of the system even in nonergodic regimes, where self-averaging might not work. For instance, both the variance `shoot-up' at short times and the enhanced time-averaged spreading of the distribution observed in Figs.~\ref{fig:figure2} and \ref{fig:figure_qkt_2} are only noticeable beyond the chaotic regime.

The present discussion thus shows that while OTOCs allow access to some aspects of the operator distribution, the quantitative connection between the two is not straightforward in practice and might become hard to probe even at the level of the first moments. It is thus desirable to think about other potential methods to probe the distribution more directly, a subject which has attracted considerable attention recently \cite{qi2019,schuster2022} and which will be the focus of an upcoming work by the authors \cite{Blocher2022NOTOC}.

\section{Conclusions and future work}
\label{sec:conclusions_and_future_work}
In this work we studied scrambling in quantum systems by analyzing the spreading of initially simple operators on a coarse-grained basis, a process which we describe via the operator distribution $\{P_k(t)\}$. We considered systems of spin-$1/2$ particles (qubits) in the basis of Pauli operators ordered by size, and kicked collective spin systems in the basis of spherical tensor operators ordered by rank.

We presented a numerical analysis of two paradigmatic models of quantum chaos in both the many-body and few-body setting: the `tilted-field' Ising model and the quantum kicked top. For both cases we computed the evolution of the operator distribution and studied its properties via computing standard distribution measures such as the mean, variance, and localization. Focusing on long-time properties, we showed that in the chaotic regimes both models evolve to spread-out distributions whose properties match the predictions for Haar-random evolution. In particular, the mean operator size (rank) in these cases is proportional to the system size $N$ while temporal fluctuations are suppressed with increasing system size, and thus the dynamics essentially equilibrates to the random distribution. In the different nonergodic regimes of these models the behavior becomes nongeneric as expected. However, several interesting unifying features are observed, like the enhancement of temporal fluctuations and the increase of the distribution variance to values above the random prediction. In the trivially integrable regimes, the distributions remain localized and show long-lived oscillations. In all the studied models we have seen that the long-time properties of the operator distribution allows one to reconstruct the integrability-to-chaos transition in the studied models. Chaotic regimes are characterized by i) the distribution mean and variance matching with Haar-random predictions and ii) the suppression of temporal oscillations. We have found that deviations from one of these two conditions indicate some degree of nonergodicity. 

We also applied the operator distribution framework to a random circuit model and showed how the different properties of the operator distribution change as a Clifford circuit is turned into a universal circuit containing non-Clifford gates. Finally, we studied the connection of the different properties of the distribution to averages of out-of-time-ordered correlators (OTOCs). The ideas and results presented in this work build on previous works and set a path where scrambling could be studied directly from the operator distribution, which can be more physically transparent than OTOCs. An important challenge is to devise experimental protocols that allow one to probe these distributions without resorting to the experimentally challenging OTOCs. \correc{Another important aspect that we leave for future work is the study of the short-time behavior of the operator distribution. In several cases, it has been observed that single-site OTOCs show exponential decay, and the corresponding exponent is associated with a quantum Lyapunov exponent. The connection discussed in Sec.~VI entails that, if all single-site OTOCs behave in this way (or rather, if the average single-site OTOC does so), then we expect an exponential \textit{increase} of the mean operator size. Interestingly, this does not say anything about the behavior of other properties of the operator distribution, and it is in principle possible to devise models in which the timescale associated with, e.g. the distribution variance is different than that given by the Lyapunov exponent}. 

Along these lines, an important path forward is to study how to apply the picture presented in Fig.~\ref{fig:figure1} to systems beyond spin-$1/2$ particles. The analysis of collective spin models studied in this work represents an advance in this direction, since these collective models are completely equivalent to single particles of a fixed total spin $J$. There is thus a notion of scrambling in a single particle for any $J>1/2$ , similar to the manner in which scrambling can be defined for a single bosonic mode \cite{zhuang2019scrambling} (note that the maximum rank for $J=1/2$ is $N=1$ and so the operator distribution is trivial). When considering, for instance, chains of spin-$1$ particles (i.e. circuits on qutrits \cite{blok2021}) or systems of interacting bosons in lattices \cite{bohrdt2017}, one should think about defining a coarse-graining of the operator basis that takes into account operator spreading within one subsystem, as well as among different subsystems.

Finally, the analysis presented in this work highlights that interesting features about the nonergodic regimes of many-body systems may be studied from the operator distribution. While this work has studied a system which is integrable by mapping to noninteracting particles, there are other systems which are integrable by other mechanisms, like the Heisenberg model \cite{santos2012}. More generally, some many-body systems show dynamical phase transitions defined from their out-of-equilibrium properties \cite{heyl2018review}. An exciting path forward is to elucidate whether some of these dynamical transitions can be viewed as a transition in the operator distribution, which in turn is initial-state independent. 

\acknowledgements
This material is based upon work supported by the U.S. Department of Energy, Office of Science, National Quantum Information Science Research Centers, Quantum Systems Accelerator (QSA). Additional support is acknowledged from Ministère de l’Économie et de l’Innovation du Québec and the Natural Sciences and Engineering Research Council of Canada. P.D.B. acknowledges support from the U.S. National Science Foundation through the FRHTP Grant No. PHY-2116246. 
\clearpage
\appendix
\section{Operator distributions for Haar-random evolution}\label{appendix:haar}
Consider the evolution of operator $\hat{O}$ given by
\begin{equation}
    \hat{O}(t) \rightarrow \hat{U}^\dagger \hat{O} \hat{U}=\sum\limits_j f\left[\hat{\Lambda}_j;\hat{U}^\dagger \hat{O}\hat{U}\right]\hat{\Lambda}_j
\end{equation} 
where $\hat{U}$ is taken from the uniform Haar distribution in SU$(d)$. We are interested in 
\begin{equation}
    \mathbb{E}\left[\left\lvert f\left[\hat{\Lambda}_j;\hat{U}^\dagger \hat{O}\hat{U}\right]\right\rvert^2\right] = \mathbb{E}\left[ \left\lvert\trace{\hat{U}^\dagger \hat{O}\hat{U}\hat{\Lambda}_j}\right\rvert^2\right],
    \label{eq:app_Expharr}
\end{equation}
where $\mathbb{E}[\cdot]$ indicates the average over the Haar measure. The result of this averaging should be independent of $j$, and thus from Eq.~(\ref{eqn:Pk_general}) we have
\begin{equation}
    \mathbb{E}[P_k]=\frac{\dim({C_k})}{\tr{\hat{O}^2}}\mathbb{E}\left[ \left\lvert\trace{\hat{U}^\dagger \hat{O}\hat{U}\hat{\Lambda}_j}\right\rvert^2\right].
\end{equation}

The evaluation of Eq.~(\ref{eq:app_Expharr}) can be performed directly by first noting that
\begin{equation}
    \trace{\hat{U}^\dagger \hat{O}\hat{U}\hat{\Lambda}_j}=\sum\limits_{lmnr}U_{nr}U^*_{ml}O_{mn}\Lambda^{(j)}_{rl}.
\end{equation}
Applying standard techniques (see Ref.~\cite{collins2006}) to integrate a degree-2 monomial in the elements $\{U_{ij}\}$, we get
\begin{widetext}
\begin{equation}
    \mathbb{E}\left[ \left\lvert\trace{\hat{U}^\dagger \hat{O}\hat{U}\hat{\Lambda}_j}\right\rvert^2\right]  =\frac{1}{d^2-1}\left\{ \trace{\hat{O}^2} \left(\trace{\hat{\Lambda}_j\hat{\Lambda}_j^\dagger} - \frac{1}{d}\left\lvert \trace{\hat{\Lambda}_j}\right\rvert^2 \right) + \trace{\hat{O}}^2 \left(\left\lvert \trace{\hat{\Lambda}_j}\right\rvert^2 - \frac{1}{d}\trace{\hat{\Lambda}_j\hat{\Lambda}_j^\dagger}\right) \right\}.
\end{equation}
\end{widetext}

Using the orthonormality of the operator basis and the fact that $\hat{O}$ is traceless, we then arrive to the simpler result
\begin{equation}
    \mathbb{E}[P_k]=\frac{\dim({C_k})}{d^2-1}.
\end{equation}
For systems of spin-$1/2$ particles, we have $d= 2^N$ and 
\begin{equation}
    \mathrm{dim}\left(C_k\right)=\binom{N}{k}3^k.
\end{equation}
Then,
\begin{equation}
    \mu = \overline{k} = \frac{3}{4}N \frac{d^2}{d^2-1}\sim \frac{3}{4}N,
\end{equation}
\noindent and 
\begin{equation}
    \overline{k^2} = \frac{3}{16}N(3N+1)\frac{d^2}{d^2-1},
\end{equation}
\noindent leading to
\begin{equation}
    \sigma^2 \simeq \frac{3}{16}N.
\end{equation}
    
For collective spin models $d=N+1$ and $\dim C_k = 2k+1$. We then get
\begin{align}
    \overline{k} &= \frac{1}{6}(4N+5)\frac{N+1}{N+2}\\
    \overline{k^2} &= \frac{1}{6}\left(N(3N+5)+1\right)\frac{N+1}{N+2},
\end{align}
\noindent which lead to $\sigma^2\sim N^2/18$. 

The calculation of the averaged IPR requires knowing $\mathbb{E}[P_k^2]$. This can be obtained by treating $P_k$ as coming from a Porter-Thomas distribution in a space of dimension $D=d^2-1$ \cite{boixo2018}, in which case $\mathbb{E}[P_k^2]=2/D(D+1)$. We then have that
\begin{equation}
    \mathbb{\eta_{\mathrm{IPR}}}=\frac{1}{(d^2-1)^2}\sum\limits_j \dim(C_j)^2 + \frac{d^2-2}{(d^2-1)d^2}
\end{equation}
Neglecting the second term, we obtain for the Pauli case ($d=2^N$)
\begin{equation}
    \mathbb{\eta_{\mathrm{IPR}}} = \frac{1}{(d^2-1)^2}\left( {}_2 F_1\left(-N,-N;1;9\right)-1\right)\sim (2.35N)^{-1/2},
\end{equation}
\noindent where ${}_2 F_1$ is a hypergeometric function and the scaling was obtained \correc{numerically}. For collective spins ($d=N+1$), we have
\begin{equation}
    \mathbb{\eta_{\mathrm{IPR}}} = \frac{1}{3}\frac{4d^3-d-3}{(d^2-1)^2}\sim \frac{4}{3N}.
\end{equation}

\section{Averaged level spacing ratio as quantum chaos indicator} \label{appendix:avg_ratio}
The average adjacent level spacing ratio measures correlations in the eigenspectrum of hermitian or unitary operators and is routinely taken as a standard measure of quantum chaos \cite{atas2013}. Given a set of eigenvalues $\{e_j\}_{j=1,\ldots,d}$ (if considering a unitary, take the real phases $\phi_j$ associated with each eigenvalue $e^{i\phi_j}$), the average adjacent spacing ratio is defined as 
\begin{equation}
    \overline{r} = \frac{1}{d}\sum\limits_{j=1}^{d-2} r_j,\ \mathrm{where}\  r_j = \frac{\max(s_j,s_{j+1})}{\min(s_j,s_{j+1})},
\end{equation}
where $s_j = e_{j+1}-e_j$. In chaotic systems, level spacing distributions $\{s_j\}$ show level repulsion following predictions from random matrix theory (RMT), and thus $\overline{r}$ takes specific values depending on the appropriate RMT ensemble. For the case of the Ising model this is the Gaussian orthogonal ensemble (GOE) and $\overline{r}_{\mathrm{GOE}}\simeq0.535$ \cite{atas2013}. For the QKT considered in this work, the appropiate ensemble is the circular unitary ensemble (CUE) for which $\overline{r}_{\mathrm{CUE}}\simeq 0.599$. Regular (integrable) systems have spectra in which the eigenvalues tend to be uncorrelated, and so the spacing distribution is instead Poissonian \cite{Haake1987}, with associated $\overline{r}_{\mathrm{POI}}=0.386$ \cite{atas2013}. As a normalized measure of chaos, we define the normalized quantity
\begin{equation}
    \overline{r}_{\mathrm{norm}} = \frac{\overline{r} - \overline{r}_{\mathrm{POI}}}{\overline{r}_{\mathrm{RMT}}-\overline{r}_{\mathrm{POI}}},
\end{equation}
where RMT corresponds to GOE or CUE depending on the system under study. The normalized measure then approaches 1 in the chaotic regime, and 0 in the nonchaotic regime. 

\section{Additional results on Ising and QKT models}\label{appendix:additional}

\begin{figure}[t!]
    \centering
    \includegraphics[width=\linewidth ]{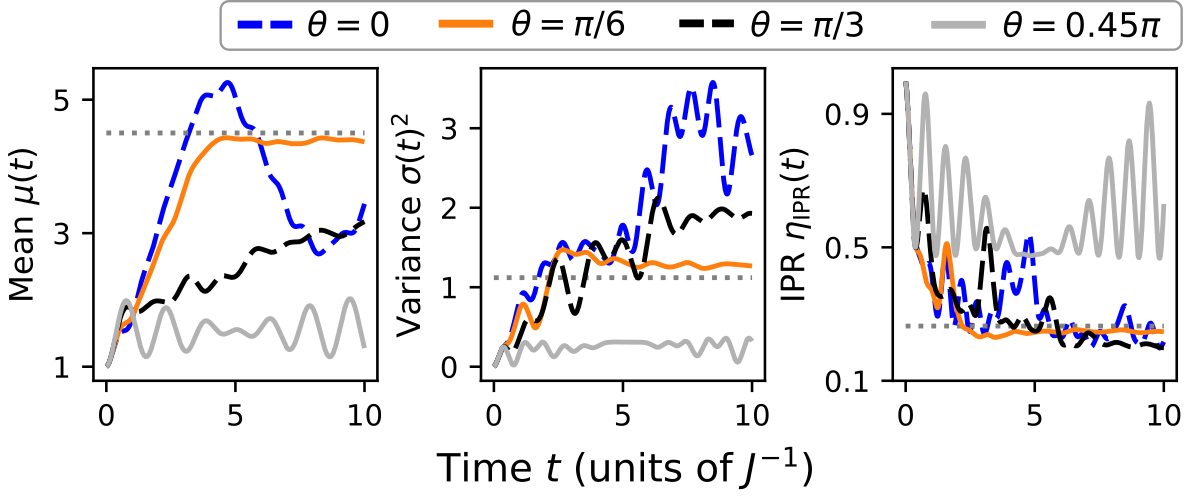}
    \caption{Measures of the probability distribution $\{P_k\}$ for the tilted field Ising model with $N=6$ and $B/J=1$ and $\hat{O}(0)=\hat{\sigma}_y^{(1)}/\sqrt{2}$}
    \label{fig:figure_app_extra}
\end{figure}

In Sec.~\ref{sec:scrambling_and_chaos_in_the_tilted_field_Ising_model} we introduced the analysis of different measures of the operator probability distribution, like the mean $\mu(t)$, variance $\sigma^2(t)$, and IPR $\eta_{\mathrm{IPR}}(t)$. In Fig.~\ref{fig:figure3} we presented the evolution of these quantities for the Ising model in the case where the initial operator sits in the middle of the chain. In Fig.~\ref{fig:figure_app_extra} we display the evolution of these properties for the situation where initial operator sits at the edge of the chain.

\begin{figure}[t!]
    \centering
    \includegraphics[width=0.9\linewidth ]{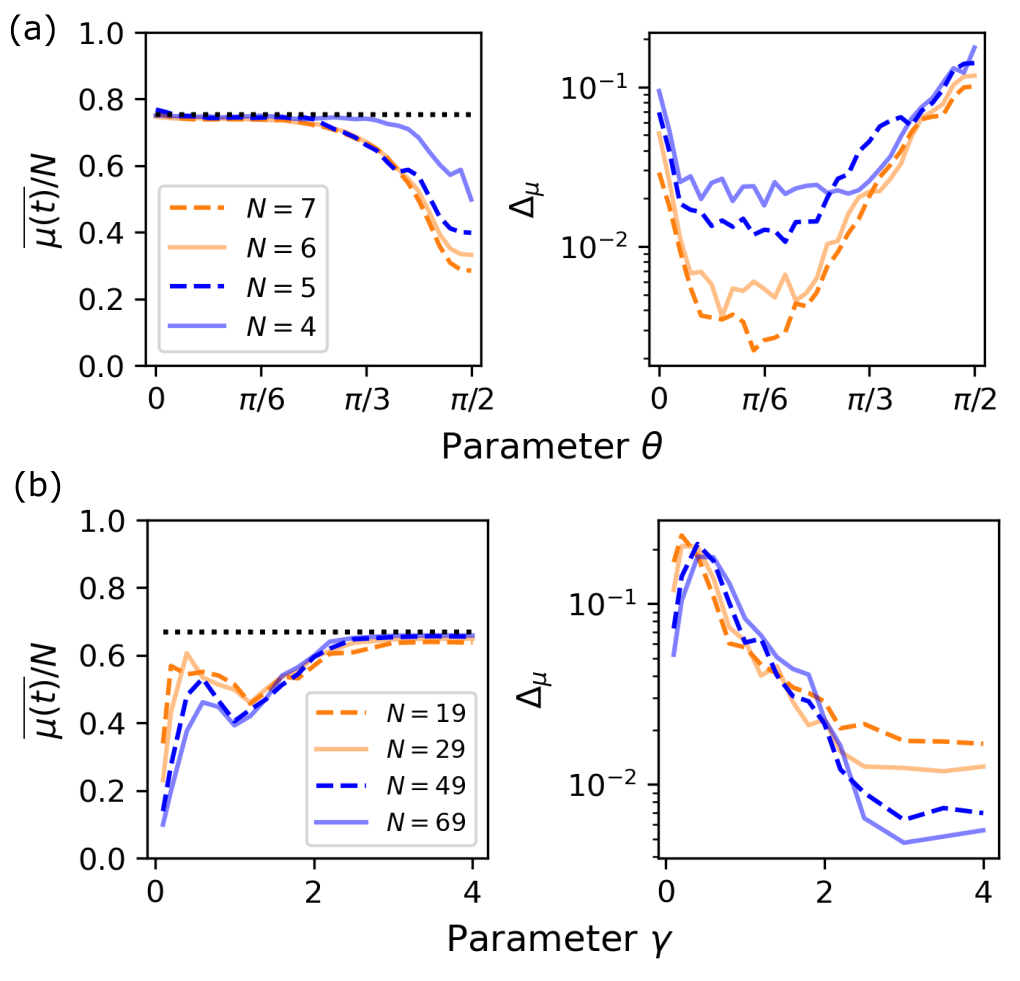}
    \caption{System size analysis for the time-averaged mean $\overline{\mu(t)}$ and associated temporal fluctuations $\Delta_\mu$ for both the Ising (a) and QKT (b) models considered in the main text. Quantities are plotted as a function of the relevant parameters for each model: the angle $\theta$ between the longitudinal and transverse field for the Ising model (a) and the nonlinearity strength parameter $\gamma$ for the QKT (b). \correc{Dotted} lines indicate the predictions from Haar-random evolution, cf. Table~\ref{tab:Pk_properties} and Appendix~\ref{appendix:haar}.}
    \label{fig:figure_app_N}
\end{figure}

In the main text we developed our analysis of the Ising and QKT models with fixed systems sizes of $N=6$ and $N=50$ respectively. Here we show additional numerical results showing that those results are representative of other cases. In Fig.~\ref{fig:figure_app_N} we illustrate the time-averaged mean of the distribution $\overline{\mu(t)}$ normalized by the system size for both models and different choices of $N$, together with the associated temporal fluctuations $\Delta_{\mu}$ (computed over the normalized measure $\overline{\mu}(t)/N$). We observe that for sufficiently large system size the normalized time-averages become independent of $N$ in the chaotic regime, as expected from the predictions of Table~\ref{tab:Pk_properties}. The temporal fluctuations approach zero in this regime for both models, and actually decrease with increasing system size $N$. Away from the chaotic regime, the behavior is quite different: the time-averaged mean shows deviations (albeit small for large $N$), and the fluctuations are actually become system-size-independent.

\section{Proofs related to the connection between OTOCs and moments of the operator distribution}\label{appendix:otoc}

In Sec.~\ref{sec:otocs_and_connections} we postulated that the average of OTOCs over the subspace $C_n$ of $n$-body Pauli operators $\mathcal{M}_n(t)$, as defined in \eqref{eq:avg_otocs}, may be written as a linear combination of moments $\mu_i(t)$ of the probability distribution up to $i=n$ on the form of \eqref{eq:OTOCConnectionMn}. In this appendix we prove this postulate, thus highlighting the connection between averages over OTOCs and moments of the operator distribution.

We start by inserting the operator expansion \eqref{eqn:evol_O} into the expression for the average of OTOCs \eqref{eq:avg_otocs} to obtain
\begin{align}
\mathcal{M}_n(t) =& \frac{1}{\text{dim}(C_n)}\sum_{\hat{Q}} \vert f[\hat{Q};\hat{W}(t)]\vert^2 \sum_{\hat{R}\in C_n}\frac{1}{d}\text{Tr}( \hat{Q}\hat{R}\hat{Q}\hat{R}) \nonumber\\
=&\frac{1}{\text{dim}(C_n)}\sum_{\hat{Q}} \vert f[\hat{Q};\hat{W}(t)]\vert^2 \left(\sum_{\hat{R}\in C_n} e^{i \phi_{\hat{Q}\hat{R}}}\right) \label{eq:ConjectureKAverageOTOC},
\end{align}
where in the first equality we have used that $\text{Tr}(\hat{Q} \hat{R} \hat{Q}^\prime \hat{R}) = 0$ unless $\hat{Q} = \hat{Q}^\prime$, and in the second equality that $\hat{Q}\hat{R} = e^{i \phi_{\hat{Q}\hat{R}}} \hat{R}\hat{Q}$. The phase factor $e^{i \phi_{\hat{Q}\hat{R}}}$ may be unpacked by writing the operator product $\hat{Q}\hat{R}$ as
\begin{equation}
    \hat{Q}\hat{R} = \bigotimes_{i=1}^N q_i r_i, \text{  where } \hat{q}_i, \hat{r}_i \in \{\mathbbm{1}, X, Y, Z\}.
\end{equation}
We note that $\hat{q}_i$ and $\hat{r}_i$ commute if $\hat{q}_i = \hat{r}_i$ or if $\hat{q}_i = \mathbbm{1}$ or $\hat{r}_i = \mathbbm{1}$, and otherwise anticommute. Thus $\hat{Q}\hat{R} = \pm \hat{R}\hat{Q} = e^{i \phi_{\hat{Q}\hat{R}}} \hat{R}\hat{Q}$ and so $\phi_{\hat{Q}\hat{R}} \in \{0,\pi\}$.

From \eqref{eq:ConjectureKAverageOTOC} we note that moments of the probability distribution $\mu_i$ appear in \eqref{eq:OTOCConnectionMn} due to the corresponding powers $k^i$ of $k = s(\hat{Q})$ appearing in the sum $\sum_{\hat{R}\in C_n} e^{i \phi_{\hat{Q}\hat{R}}}$ of \eqref{eq:ConjectureKAverageOTOC}. To prove the validity of \eqref{eq:OTOCConnectionMn} we thus show in the following that for any $n\in\mathbb{N}$, $n \leq N$ the sum $\sum_{\hat{R}\in C_n} e^{i \phi_{\hat{Q}\hat{R}}}$ contains powers of $k = s(\hat{Q})$ up to (and including) the $n$th power.

Let $n\in\mathbbm{N}$, $n\leq N$ be given. For any $\hat{R} \in C_n$ we have $n = s(\hat{R})$ sites labeled $i_1$, $i_2$, \ldots, $i_n$ on which $\hat{r}_{i_j} \neq \mathbbm{1}$ ($j = 1,2,\ldots,n$), while the remaining $N-s(\hat{R})$ sites are identities $\hat{r}_{i_j} = \mathbbm{1}$ (for $j = n+1,\ldots,N$). For a given operator $\hat{Q}$, there are now $n+1$ possible cases that occur as we sum over $\hat{R} \in C_n$:
\begin{enumerate}
    \item[0.] $\hat{q}_{i_j} = \mathbbm{1}$ for all $j = 1,2,\ldots,n$.
    \item One $\hat{q}_{i_j} \neq \mathbbm{1}$ for some $j_1$, and the remaining $\hat{q}_{i_j} = \mathbbm{1}$ for $j\neq j_1$, $j \leq n$.
    \item Two $\hat{q}_{i_j} \neq \mathbbm{1}$ for some $(j_1,j_2)$, and the remaining $\hat{q}_{i_j} = \mathbbm{1}$ for $j\neq j_1,j_2$, $j \leq n$.
    \item[\vdots]
    \item[n.] $\hat{q}_{i_j} \neq \mathbbm{1}$ for all $j = 1,2,\ldots,n$.
\end{enumerate}
We let $m$ be the number of non-identity $\hat{q}_{i_j}$ in a given case, which also serves to label the above $n+1$ cases. For each of these cases we need to determine their occurrence $O_m^n(\hat{Q})$ and value $V_m^n(\hat{Q})$ such that we may calculate $\sum_{\hat{R}\in C_n} e^{i \phi_{\hat{Q}\hat{R}}} = \sum_m O_m^n(\hat{Q}) V_m^n(\hat{Q})$.

The occurrence of each case is straightforward to determine as
\begin{equation}
    O_m^n(\hat{Q}) = \begin{pmatrix} N-s(\hat{Q}) \\ n-m \end{pmatrix} \begin{pmatrix} s(\hat{Q}) \\ m \end{pmatrix}, \label{eq:appD_combinatorics}
\end{equation}
where the first binomial coefficient is the number of unique ways to choose the sites on which $\hat{q}_{i_j} = \mathbbm{1}$, whereas the second binomial coefficient is the number of unique ways to choose the non-identity sites. For the value $V_m^n(\hat{Q})$ of a given case $m$, each site with $\hat{q}_{i_j} = \mathbbm{1}$ simply yields a factor $3$ to the number of outcomes $e^{i \phi_{\hat{Q}\hat{R}}} = +1$ and $e^{i \phi_{\hat{Q}\hat{R}}} = -1$ (for a total factor $3^{n-m}$). The $m$ non-identity sites yield $e^{i \phi_{\hat{Q}\hat{R}}} = +1$ only if the number of sites for which $\hat{q}_{i_j} \neq \hat{r}_{i_j}$ is even. We are thus looking for the number of pairs, quadruplets, sextuplets, and higher order even tuplets of sites that one can create. Pairs provide $2^2 = 4$ different combinations of operators on the two sites, quadruplets $2^4 = 16$ different combinations, and so on. The remaining combinations must yield $e^{i \phi_{\hat{Q}\hat{R}}} = -1$, and as there are $3^n$ different combinations of operators, the value of $V_m^n(\hat{Q})$ for case $m$ takes the form
\begin{equation}
    V_m^n(\hat{Q}) =  -3^n + 2 \cdot 3^{n-m} \sum_{i=0}^{\text{floor}(m/2)} \begin{pmatrix}m \\ 2i \end{pmatrix} 2^{2i}
\end{equation}
which only depends on the number of non-identity sites $m$ and not on the operator $\hat{Q}$. We may thus omit the explicit dependence on $\hat{Q}$ and write $V_m^n$.

\begin{figure}
    \centering
    \includegraphics[width=0.8\linewidth]{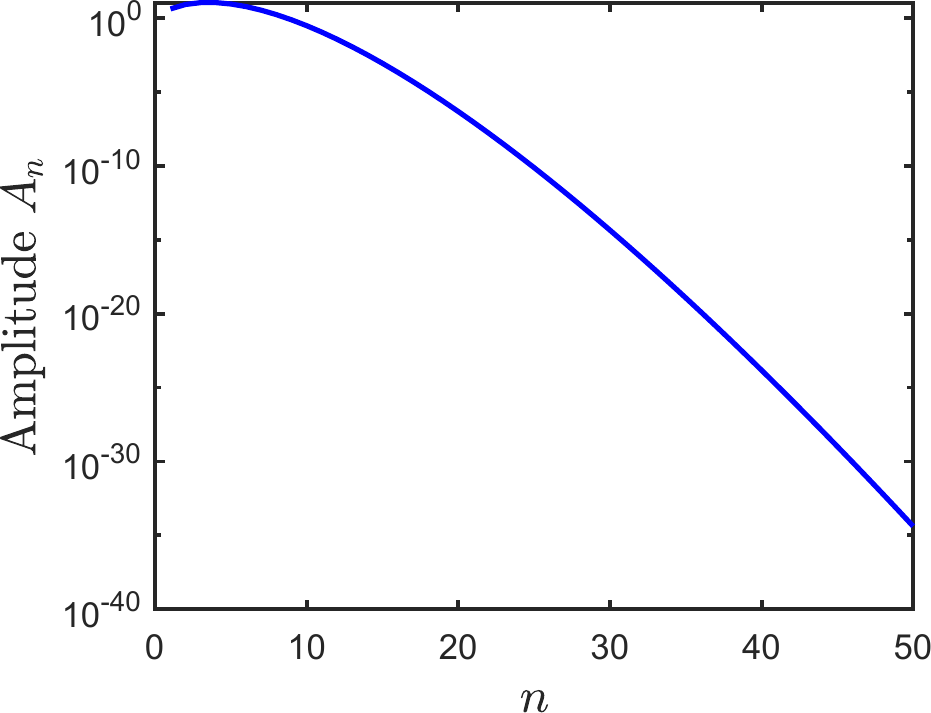}
    \caption{The amplitude $A_n$ of the $s(\hat{Q})^n$ term, as defined in \eqref{eq:ConjectureAmplitude}, visualized as a function of $n$. We observe that the amplitude is non-zero for all $n$ considered here, albeit exponentially decreasing. This demonstrates numerically that the polynomial \eqref{eq:ConjecturePolynomial} is indeed order $n$ as desired for $n =1,2,\ldots,50$.}
    \label{fig:AmplitudeSk}
\end{figure}

We are now ready to evaluate the sum $\sum_{\hat{R}\in C_n} e^{i \phi_{\hat{Q}\hat{R}}} = \sum_m O_m^n(\hat{Q}) V_m^n$. We immediately note that $O_m^n(\hat{Q})$ contains all powers $s(\hat{Q})^j$ up to $j=n$ due to the product of the two binomial coefficients in \eqref{eq:appD_combinatorics}, and hence $\sum_{\hat{R}\in C_n} e^{i \phi_{\hat{Q}\hat{R}}}$ is a sum over different $n$th order polynomials in $s(\hat{Q})$ with coefficients $V_m^n$. Since the sum of two $n$th order polynomials is at most $n$th order itself, we conclude that
\begin{equation}
    \sum_{\hat{R}\in C_n} e^{i \phi_{\hat{Q}\hat{R}}} = \text{polynomial in $s(\hat{Q})$ of order $\leq n$}. \label{eq:ConjecturePolynomial}
\end{equation}
To establish the result of \eqref{eq:OTOCConnectionMn}, we need to show that amplitude of the $n$th order term in the polynomial is non-zero. The amplitude reads
\begin{equation}
    A_n = \sum_{m=0}^n \frac{(-1)^{n-m} V_m^n}{(n-m)!\, m!}, \label{eq:ConjectureAmplitude}
\end{equation}
where the alternating sign and the denominator are due to the form of \eqref{eq:appD_combinatorics}. Although we have not been successful in showing that this amplitude is non-zero for all $n$, \eqref{eq:ConjectureAmplitude} is readily evaluated numerically for $n$ of modest size, limited only by the two factorials in the denominator. In Fig.~(\ref{fig:AmplitudeSk}) the absolute value of the amplitude is displayed for $1 \leq n \leq 50$, and we see that the amplitude, although decreasing super-exponentially, is non-zero for all $n$s considered. The non-vanishing amplitude of the $s(\hat{Q})^n$ term ensures that the average of OTOCs over the subspace $C_n$ of Pauli operators is a linear combination of all moments up to and including the $n$th moment $\braket{s(t)^n}$, confirming \eqref{eq:appD_combinatorics} for all $n$ that we have been able to access numerically.

\bibliography{reference.bib}
\end{document}